\documentclass{article}

\usepackage{microtype}
\usepackage{graphicx}
\usepackage{subfigure}
\usepackage{booktabs} 

\usepackage{hyperref}

\usepackage[accepted]{icml2024}

\usepackage{amsmath}
\usepackage{amssymb}
\usepackage{mathtools}
\usepackage{amsthm}

\DeclareMathOperator{\bin}{bin}

\usepackage[capitalize,noabbrev]{cleveref}

\theoremstyle{plain}

\theoremstyle{definition}

\theoremstyle{remark}

\usepackage{hyperref}
\usepackage{url}
\usepackage{nicefrac}       
\usepackage{amssymb}
\usepackage{amsmath}
\usepackage{multirow}
\usepackage{subcaption}
\usepackage{float}
\usepackage{mathtools}
\usepackage{bbold}
\usepackage{subcaption}
\usepackage{caption} 
\usepackage{graphicx}
\usepackage{booktabs} 
\usepackage{amssymb}
\usepackage{amsmath}
\usepackage{multirow}
\usepackage{subcaption}
\usepackage{float}
\usepackage{mathtools}
\usepackage{bbold}
\usepackage{subcaption}
\usepackage{graphicx}
\usepackage{multicol}
\usepackage{makecell}
\usepackage{bbm}

\usepackage[textsize=tiny]{todonotes}

\icmltitlerunning{Learning Linear Block Error Correction Codes}

\begin{document}

\twocolumn[
\icmltitle{Learning Linear Block Error Correction Codes}



\icmlsetsymbol{equal}{*}

\begin{icmlauthorlist}
\icmlauthor{Yoni Choukroun}{comp}
\icmlauthor{Lior Wolf}{comp}
\end{icmlauthorlist}
\icmlaffiliation{comp}{The Blavatnik School of Computer Science, Tel Aviv University}


\icmlcorrespondingauthor{Yoni Choukroun}{choukroun.yoni@gmail.com}
\icmlcorrespondingauthor{Lior Wolf}{wolf@cs.tau.ac.il}

\icmlkeywords{Machine Learning, ICML}

\vskip 0.3in
]



\printAffiliationsAndNotice{}  

\begin{abstract}
Error correction codes are a crucial part of the physical communication layer, ensuring the reliable transfer of data over noisy channels.
The design of optimal linear block codes capable of being efficiently decoded is of major concern, especially for short block lengths. 
While neural decoders have recently demonstrated their advantage over classical decoding techniques, the neural design of the codes remains a challenge. 
In this work, we propose for the first time a unified encoder-decoder training of binary linear block codes.
To this end, we adapt the coding setting to support efficient and differentiable training of the code for end-to-end optimization over the order two Galois field. 
We also propose a novel Transformer model in which the self-attention masking is performed in a differentiable fashion for the efficient backpropagation of the code gradient. Our results show that 
(i) the proposed decoder outperforms existing neural decoding on conventional codes, (ii) the suggested framework generates codes that outperform the {analogous} conventional codes, and (iii) the codes we developed not only excel with our decoder but also show enhanced performance with traditional decoding techniques.
\end{abstract}
\section{Introduction}
In the modern era of information technology, maintaining strong and reliable communication despite interference in transmission channels is a significant priority. 
It necessitates the development of codes designed for resilient transmission over noisy channels. 
Among the existing family of codes, linear block codes represent a widely used class of error correction codes (ECC), which benefits from decades of research and design. 
Neural methods \cite{nachmani2016learning,gruber2017deep,bennatan2018deep,choukroun2022error} have been applied as successful heuristics to the NP-hard maximum likelihood (ML) problem of decoding. 
However, creating high-performance codes in conjunction with their decoders remains a challenge, particularly in the contemporary realm of finite-length codes.

Linear block codes are generally designed upon asymptotical analysis based on mathematical principles or upon pseudorandom generation. 
The Shannon channel capacity theorem \cite{shannon1948mathematical} demonstrated first the existence of a coding technique that allows an arbitrarily small error probability under maximum likelihood decoding when the transmission rate remains closely below the channel capacity.
Then, the capacity can be asymptotically achieved using random linear block codes, encoded in polynomial time. Polar codes \cite{arikan2008channel} are a family of codes proven to be capacity-achieving under successive cancellation (SC) decoding with an explicit construction method based on recursive channel combination. 
Low-Density Parity-Check (LDPC) codes \cite{gallager1962low} are provably close to the capacity generally obtained from random sparse bipartite graphs with appropriate degree distribution. 
LDPC codes are effectively decoded via the efficient Belief Propagation algorithm \cite{pearl1988probabilistic}. 

While neural methods have been applied successfully to train decoders of existing codes, attempts to use machine learning for code design are far fewer. 
There have been attempts to learn non-linear {continuous} codes along with their neural decoders \cite{AutoencoderComm, jiang2019turbo}, but the very high degree of non-differentiability of the ubiquitous \emph{binary linear block} codes makes their design a major challenge.

Our contributions include (i) showing for the first time that it is possible to directly optimize binary linear block codes along with a neural decoder, in a unified and differentiable fashion. 

We optimize the denoising capabilities of a learned code with respect to a neural decoder that is trained jointly.  A central question then becomes whether the learned code is tailored for a given neural decoder only, or whether the framework provides a universally good code in some sense. 
 (ii) We present compelling evidence supporting that the optimized code exhibits improved performance compared to other codes of the same size, irrespective of the decoder used. This outcome is somewhat unexpected, in light of the significantly non-convex objective associated with the integrated optimization process.

To achieve these innovative results, we make multiple technical contributions: (iii) we adapt the error correction coding setting for efficient differentiable training, (iv) we solve the highly not-differentiable optimization over the binary finite field with binarization and polarization methods, and (v) we improve over the existing ECC Transformer-based models by creating a mask derived from the parity-check matrix in a differentiable manner, allowing the efficient backpropagation of the gradients with respect to the code through the neural decoder's layers.
\section{Related Works}
\textcolor{black}{Neural decoder contributions generally focus on short and moderate-length codes for two main reasons.
First, classical decoders are proven to reach the capacity of the channel for large codes, preventing any potential enhancement. 
Second, the emergence of applications driven by the Internet of Things created the requirement for optimal decoders of short to moderate-length codes. For example, 5G Polar codes have code lengths of 32 to 1024 \citep{ETSI}.}

Previous work on neural decoders is generally divided into two main classes: model-based and model-free. 
Model-based decoders implement parameterized versions of classical Belief Propagation (BP) decoders, where the Tanner graph is unfolded into an NN in which weights are assigned to each variable edge. This results in an improvement in comparison to the baseline BP method for short codes \citep{nachmani2016learning,nachmani2019hyper,raviv2020graph,raviv2023crc,kwak2023boosting}. While model-based decoders benefit from a strong theoretical background, the architecture is overly restrictive, which generally enforces its coupling with high-complexity NN \cite{nachmani2021autoregressive}. Also, the improvement gain generally vanishes for more iterations and longer codewords \cite{sionna}.

Model-free decoders employ general types of neural network architectures. Earlier approaches \citep{cammerer2017scaling,gruber2017deep,kim2018communication} employed stacked fully connected (FC) networks, convolutional neural networks (CNN) \cite{jiang2019deepturbo} or recurrent neural networks (RNNs) that have difficulties in learning the code, since no prior can be straightforwardly established. Similarly, \citep{bennatan2018deep} extended the classical syndrome decoding by employing a channel output preprocessing, which further adds the magnitude to the syndrome vector such that the decoder remains provably invariant to the codeword, avoiding overfitting the exponential number of codewords. 
Recently, several works based on the Transformer \cite{vaswani2017attention} architecture have been adapted to ECC.
\citep{choukroun2022error} first introduced the Error Correction Code Transformer (ECCT), obtaining state-of-the-art performance. Subsequently, \citep{choukroun2022zdenoising} extended the denoising diffusion paradigm to ECC, further improving the SOTA by large margins.
\citep{choukroun2023deep} employed the ECCT to syndrome decoding for quantum error correction.
Finally, \citep{choukroun2024found} proposed a foundation neural decoder, capable of decoding and generalizing to any code, length, and rate, enabling the potential deployment of a single universal neural decoder for every type of code.
At the intersection of Transformers and neural BP solutions, \citep{cammerer2022graph} proposed a graph neural network decoder built upon the Tanner graph.
Recently, and following \citep{bennatan2018deep,choukroun2022error, raviv2020graph}, \citep{park2023mask} has shown that different parity-check matrices describing the same code provide different performance on the ECCT.

While neural decoders show improved performance in various communication settings, there has been very limited success in the design of novel neural coding methods.
Most of the existing works attack the unified training design in a classical deep encoder-decoder fashion (based on FC, CNN, or RNN), where the codes and the modulations are integrated in a fully classical differentiable fashion.
\citep{AutoencoderComm} developed continuous (7,4) code with modulation power constraint matching the Hamming code performance. 
Joint designs have been proposed for feedback channels \cite{kim2018deepcode} and Turbo codes \cite{kim2018deepcode}.
End-to-end training under non-differentiable modulations has been studied in \citep{aoudia2018end,ye2018channel}

However, these methods are generally problematic, since they make use of heavy deep learning-based encoding-decoding solutions in the {continuous} domain, which is far from practical encoding and decoding deployment. Moreover \cite{jiang2019turbo}, neural codes remain far from capacity-approaching performance because of the high level of non-differentiability, as well as the difficulties in inducing the code through the neural decoder.

\section{Background}
We provide the necessary background on error correction coding and the Transformer architectures for ECC.
\subsection{Coding}
\label{sec:codingbck}
We assume a standard transmission protocol for messages $m \in \{0, 1\}^{k}$ using a linear code $C$, defined by a {generator} matrix $G\in \{0,1\}^{k \times n}$ and the parity check matrix $H\in \{0,1\}^{(n - k) \times n}$, such that $GH^{T}=0$ over the order 2 Galois field $GF(2)$.
{The parity check matrix $H$ entails what is known as a Tanner graph, which consists of $n$ variable nodes and $(n-k)$ check nodes. The edges of this graph correspond to the on-bits in each column of the matrix $H$.}

The input message $m \in \{0, 1\}^{k}$ is encoded by $G$ to a codeword $x=mG \in C \subset \{0, 1\}^{n}$ satisfying $Hx=0$ and is transmitted via a Binary-Input Symmetric-Output channel, e.g., an additive white Gaussian noise (AWGN) channel.
Let $y$ denote the channel output represented as $y=x_{s}+\varepsilon$, where $x_s$ denotes the Binary Phase Shift Keying (BPSK) modulation of $x$ (i.e., over $\{\pm 1\}$), and $\varepsilon$ is a random noise independent of the transmitted $x$. The main goal of the decoder $f:\mathbb{R}^{n}\rightarrow \mathbb{R}^{n}$ is to provide a soft approximation $\hat{x}=f(y)$ of the codeword. 

Following \citet{bennatan2018deep, choukroun2022error, choukroun2022zdenoising}, the surrogate decoder objective is defined by the prediction of the equivalent multiplicative noise $\tilde{\varepsilon}$ such that $y=x_{s}\odot \tilde{\varepsilon}$, with $\odot$ the Hadamard product. 
{Extending classical syndrome decoding,} the decoder preprocesses the channel output $y$ by concatenating the provably \emph{codeword-independent} magnitude and syndrome vectors, such that \mbox{$h(y)\coloneqq[|y|,s(y)] \in \mathbb{R}^{2n-k}$}, where $[\cdot ,\cdot]$ denotes vector/matrix concatenation and $s(y)$ denotes the syndrome defined by $s(y)= H\text{bin}(y)$ with $\text{bin}(y)$ the binary mapping of $y$.
In this case, the soft codeword prediction is given by the denoising task defined as \mbox{$\hat{x} = f(h(y))\odot y$.}

\subsection{Transformers for Error Correction Code}

The seminal Transformer architecture was first introduced as a novel, attention-based architecture for machine translation~\cite{vaswani2017attention}. The input sequence is embedded into a high-dimensional space, coupled with positional embedding for each element. The embeddings are then propagated through multiple normalized self-attention and feed-forward blocks. 
The self-attention mechanism is based on a trainable associative memory with (key, value) vector pairs, where a query vector $q \in \mathbb{R}^d$ is matched against a set of $k$ key vectors using scaled inner products, such that \mbox{$A(Q,K,V)=\text{Softmax}({d^{-\nicefrac{1}{2}}(QK^{T}}))V$}.
Here, \mbox{$Q \in \mathbb{R}^{N \times d}$}, $K \in \mathbb{R}^{k \times d}$ and $V \in \mathbb{R}^{k \times d}$ represent the stacked $N$ queries, $k$ keys and values tensors, respectively.
Keys, queries, and values are obtained using linear transformations of representations of the sequence's elements, and a multi-head self-attention scheme is deployed by extending the self-attention mechanism to multiple attention heads.

The  Error Correction Code Transformer \citep{choukroun2022error} is a state-of-the-art neural error decoder. {Its initial embedding is defined by encoding $h(y)$, viewed as a sequence of length $2n-k$ where each bit is encoded into a high-dimensional space with its own {\color{black}(position-dependent)} embedding vector.}  To integrate information about the code, a binary masking derived from the parity-check $H$ matrix is integrated into the self-attention mechanism  $A_{H}(Q,K,V)=\text{Softmax}({d^{-\nicefrac{1}{2}}(QK^{T}+ g(H)}))V$, where \mbox{$g(H):\{0,1\}^{(n-k)\times n}\rightarrow \{-\infty,0\}^{(2n-k)\times (2n-k)}$}. Specifically, the mask $g(H)$ is obtained as the adjacency matrix of the Tanner graph extended to two-ring connectivity.  Finally, the prediction module is implemented with two standard linear layers.

Recently, the Foundation ECCT (FECCT) \cite{choukroun2024found} matched the ECCT's performance while being fully code-, length-, and rate-invariant. FECCT enables one to train a single decoder on several codes and demonstrates strong generalization capabilities on unseen codes. The length invariance of the initial embedding {is obtained using a single input encoding vector for all the magnitude elements and two position-invariant embedding vectors for representing the binary syndrome elements}.
The positional embedding, as well as the code, are integrated as \emph{relative} positional encoding into the self-attention via a {parameterized soft} mapping of the node distances in the Tanner graph, in order to modulate the self-attention tensor such that $A_{H}(Q,K,V)=\big(\text{Softmax}(d^{-\nicefrac{1}{2}}{QK^{T}})\odot \psi(\mathcal{G}(H))\big)V$, where $\psi(\mathcal{G}(H))$ is the {learned mapping of the distances between the nodes of the Tanner graph. Finally, to remain code-invariant, the final prediction module is conditioned on the parity-check matrix $H$ by selecting the relevant variable nodes from the parity-check embeddings.}
\section{Method}
We present the setting of the proposed framework and the elements of the proposed neural decoder, its complete architecture, and the training procedure.

\subsection{End-to-End Optimization}

\begin{figure}[t]
\centering
\includegraphics[width=1\columnwidth]{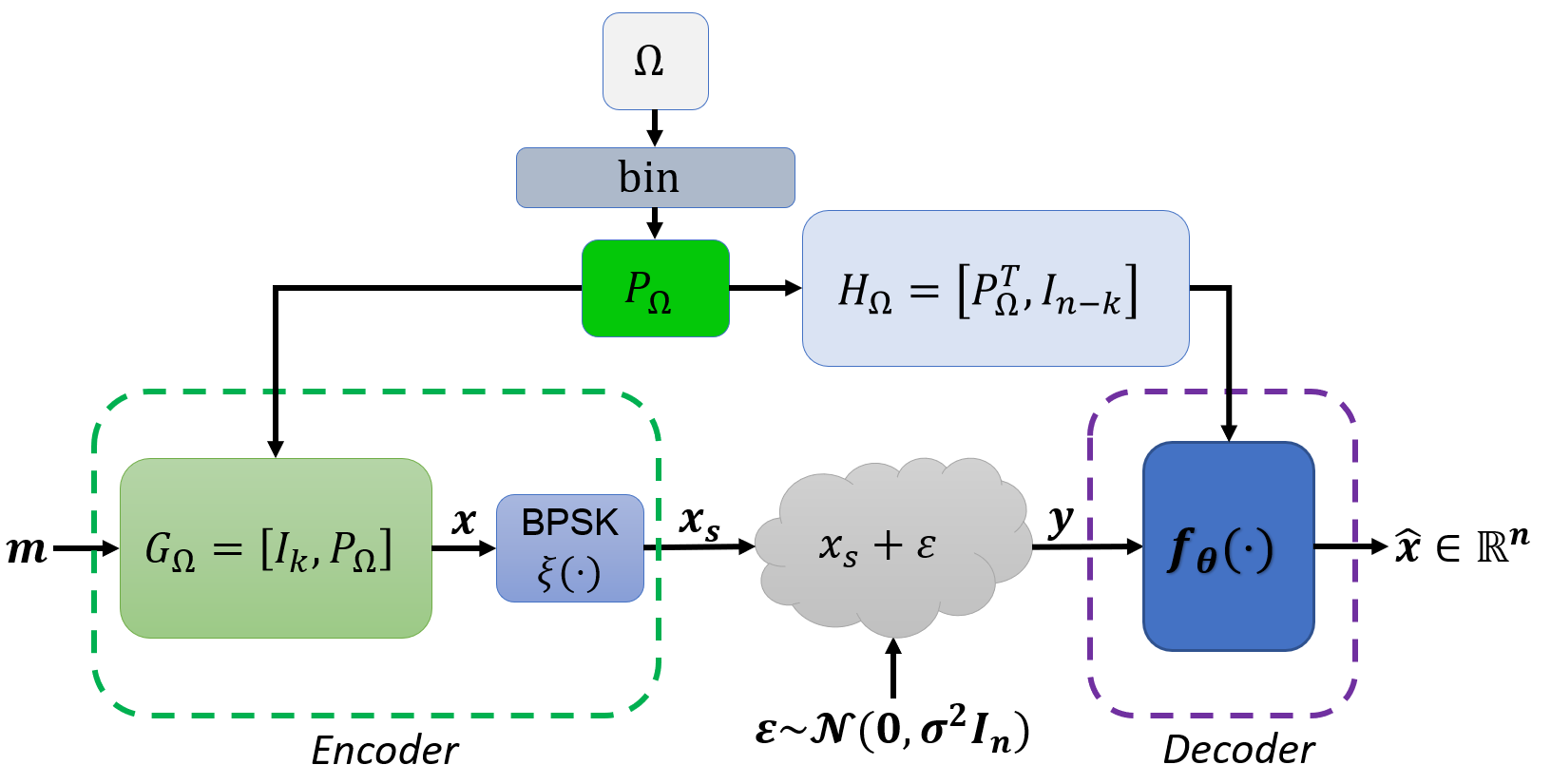}
\caption{Illustration of the proposed end-to-end communication system. Our work focuses on the unified design and co-training of the code induced by $\Omega$ and of the parameterized decoder $f_{\theta}$.}
\label{fig:ecc_illustration}
\end{figure}

We assume the {\emph{standard} (also referred to as canonical or systematic)} form of the code to ensure its efficient and differentiable optimization. In the standard form, the generator matrix is defined as $G=[I_{k},P]$ with $P\in \{0,1\}^{k\times n-k}$ and the parity check matrix is induced as $H=[P^{T},I_{n-k}]$. Using a general matrix form for $G$ would allow a greater degree of freedom in design, but the \emph{differentiable} and \emph{fast} design of the corresponding full-rank parity-check matrix defining the code kernel from $G$ would become a challenge.

To obtain trainable codes, we propose to parameterize the matrix $P$ such that
\begin{equation}
\begin{aligned}
\label{eq:p_def}
P\coloneqq P_{\Omega} = \bin(\Omega),
\end{aligned}
\end{equation}
where $\Omega \in \mathbb{R}^{k\times n-k}$ denotes the \emph{trainable} parameterized version of $P$ and $\bin:\mathbb{R}\rightarrow \{0,1\}$ denotes the point-wise binarization function.

Given a neural decoder $f_{\theta}:\mathbb{R}^{2n-k}\rightarrow \mathbb{R}^{n}$ parameterized by $\theta$, the parameterized generator matrix $G_{\Omega}=[I_{k},P_{\Omega}]$ and parity check matrix $H_{\Omega}=[P_{\Omega}^{T},I_{n-k}]$, the objective is now defined in a {unified} end-to-end encoding-decoding fashion, as opposed to the standard neural decoding optimization task {in which only the decoder is trained}.

Defining $\phi(\cdot,\cdot)$ the matrix multiplication over $GF(2)$ (i.e., modulo 2) and the bipolar mapping \mbox{\color{black}$\xi:\{0,1\}\rightarrow \{\pm 1\}$} as \mbox{$\xi(u)=1-2u, u\in \{0,1\}$} 
, the objective is given by
\begin{equation}
\begin{aligned}
\label{eq:e2e_obj}
\mathcal{L}(\Omega, \theta) 
=
\mathbb{E}_{m\sim \text{Bern}^{k}({\nicefrac{1}{2}}), \varepsilon \sim \mathcal{Z}} 
&\text{BCE}\big(f_{\theta}(h_{\Omega}(y_{\Omega})),\text{bin}(\tilde{\varepsilon})\big) 
\end{aligned}
\end{equation}
Here, $y_{\Omega}=\xi(\phi(m,G_{\Omega}))+{\varepsilon}$ denotes the parameterized channel output, $h_{\Omega}(y_{\Omega})=[|y_{\Omega}|,H_{\Omega}\text{bin}(y_{\Omega})]$ is the parameterized codeword-invariant preprocessing, $\mathcal{Z}$ denotes the distribution of the channel noise used for the training, $\text{BCE}$ denotes the binary cross entropy loss, and $\tilde{\varepsilon}$ is defined in Sec.~\ref{sec:codingbck} above.
An illustration of the proposed end-to-end communication system is given in Figure \ref{fig:ecc_illustration}.
{\color{black}Constraints on the code (e.g., sparsity, structure) can be further added to the training objective via its regularization. }

{While one could argue that the definition and optimization of the code via the decoder only (without the generator) is sufficient via its syndrome computation, we show in Appendix \ref{appendix:rep_code} that the integration of the whole encoding-decoding pipeline (i.e., $G$ and $H$) is crucial for efficient backpropagation.}

\subsection{Optimization over $GF(2)$}
The major problem in the end-to-end training objective is the use of the highly non-differentiable $\phi$ and $\text{bin}$ functions.

Here, we propose to perform the optimization of the binarization function $\bin$ via the straight-through estimator (STE) \citep{bengio2013estimating} defined such that 
\begin{equation}
\begin{aligned}
\label{eq:bin_func}
\begin{cases}
    \bin(u) = \xi^{-1}\big(\text{sign}(u)\big)\\
    \frac{\partial\text{bin}(u)}{\partial u} = -\frac{\mathbbm{1}_{|u|\leq \tau}}{2}\\
\end{cases}
\end{aligned}
\end{equation}
with $\tau$ the thresholding scalar that stops the weights $\Omega$ from growing overly large \cite{courbariaux2015binaryconnect}.

The optimization of $\phi$  is obtained using a differentiable equivalence mapping of the XOR ($\oplus$; i.e., sum over $GF(2)$) operation using the following property: \mbox{$\xi(u\oplus v)=\xi(u)\xi(v), \forall u,v\in \{0,1\}$}.
Thus, without loss of generality, with ${G_{\Omega}}_{i}$ being the $i$-th column of ${G_{\Omega}}$ and $m$ a binary vector, we have $\forall i \in \{1\dots n\}$
\begin{equation}
\label{eq:polar-eq}
\big(\phi({m,G_{\Omega}})\big)_{i} \coloneqq {G_{\Omega}}_{i}\oplus m=
\xi^{-1}\bigg(\Pi_{j=1}^{k} \xi\big((G_{\Omega})_{ij}\cdot m_{j}\big)\bigg).
\end{equation}
The new form defines a multilinear polynomial (potentially inducing saddle-point optimization) over the classical linear dot-product defined over $\mathbb{R}$ and the gradient can now be computed in a differentiable manner.

\subsection{Differentiable Masking}

The masking allows the integration of information about the code into the self-attention tensor. 
The masking derived from the Tanner graph connectivity can be soft \cite{choukroun2024found} or hard \cite{choukroun2022error} and can be placed at different locations of the self-attention computation.
However, existing masking methods induced from the code are extracted once in a non-differentiable fashion, i.e., no information can be backpropagated during the optimization from the mask to the code (i.e., the parity matrix).

In order to allow the integration of the code through the self-attention while permitting its differentiable optimization, we learn a parameterized mapping $\psi_{\gamma}: \mathbb{N}\rightarrow \mathbb{R}$ of the elements constituting the mask, which is derived by the parity-check matrix, such that
\begin{equation}
\begin{aligned}
\label{transformer_att_e2eecct}
A_{H}(Q,K,V)=\text{Softmax}\bigg(\frac{QK^{T}+\psi_{\gamma}\big(g(H_{\Omega})\big)}{\sqrt{d}}\bigg)V,
\end{aligned}
\end{equation}
where the mask $g(H_{\Omega})\in \mathbb{N}^{(2n-k)\times (2n-k)}$ is defined by

\begin{equation}
\label{eq:e2e_mask}
\begin{aligned}
g(H_{\Omega}) = 
\begin{pmatrix}
H_{\Omega}^{T}H_{\Omega} & H_{\Omega}\\
H_{\Omega}^{T} & H_{\Omega}H_{\Omega}^{T}
\end{pmatrix}
\end{aligned}
\end{equation}

Since $H_{\Omega}$ represents the bipartite Tanner graph, the mask diagonal block elements can be seen as the two-step transitioning connectivity, i.e., the number of paths of length two between every two nodes.
The $n\times n$ top-left block matrix represents the two-step transition matrix between every two variable nodes, while the $(n-k)\times (n-k)$ bottom-right block matrix represents the two-step transition matrix between every two parity-check nodes. The diagonal elements of these matrices denote the degree of each node.
Since the block off-diagonal is defined by $H_{\Omega}$ solely, it straightforwardly defines the relationship between the parity nodes and the variable nodes of the corresponding graph.

\begin{figure}[t]
\centering
\includegraphics[width=1\columnwidth]{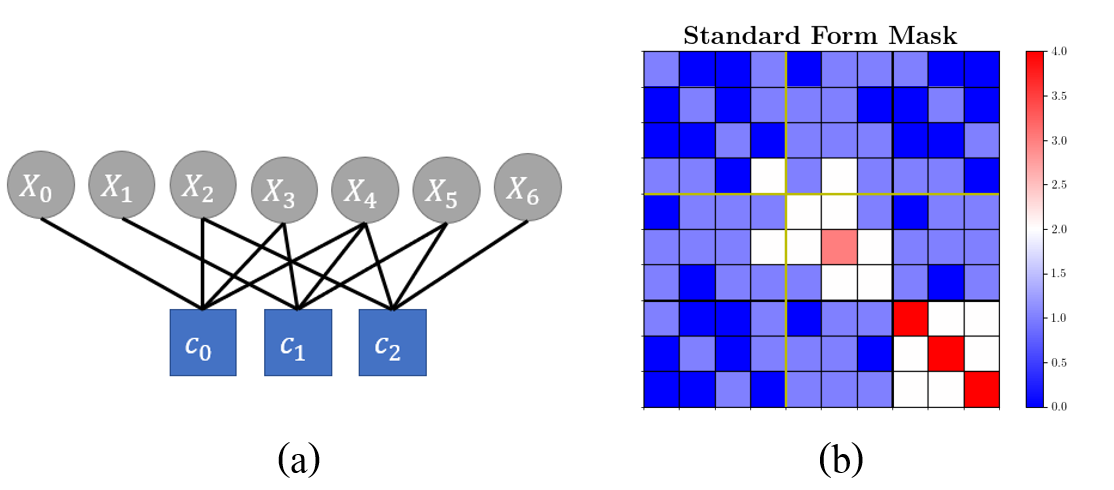}
\caption{For the Hamming(7,4) code: (a) the Tanner graph, (b) the proposed differentiable masking for the standardized version of the code.}
\label{fig:e2e_masking}
\end{figure}

This way, the gradient $\nabla_{\Omega} \mathcal{L}$ can be backpropagated through the self-attention layers along the network to provide a decoder-aware code. An illustration of the proposed masking method is given in Figure \ref{fig:e2e_masking}.

\subsection{Architecture}
\begin{figure}[t]
\centering
\includegraphics[width=1\columnwidth,]{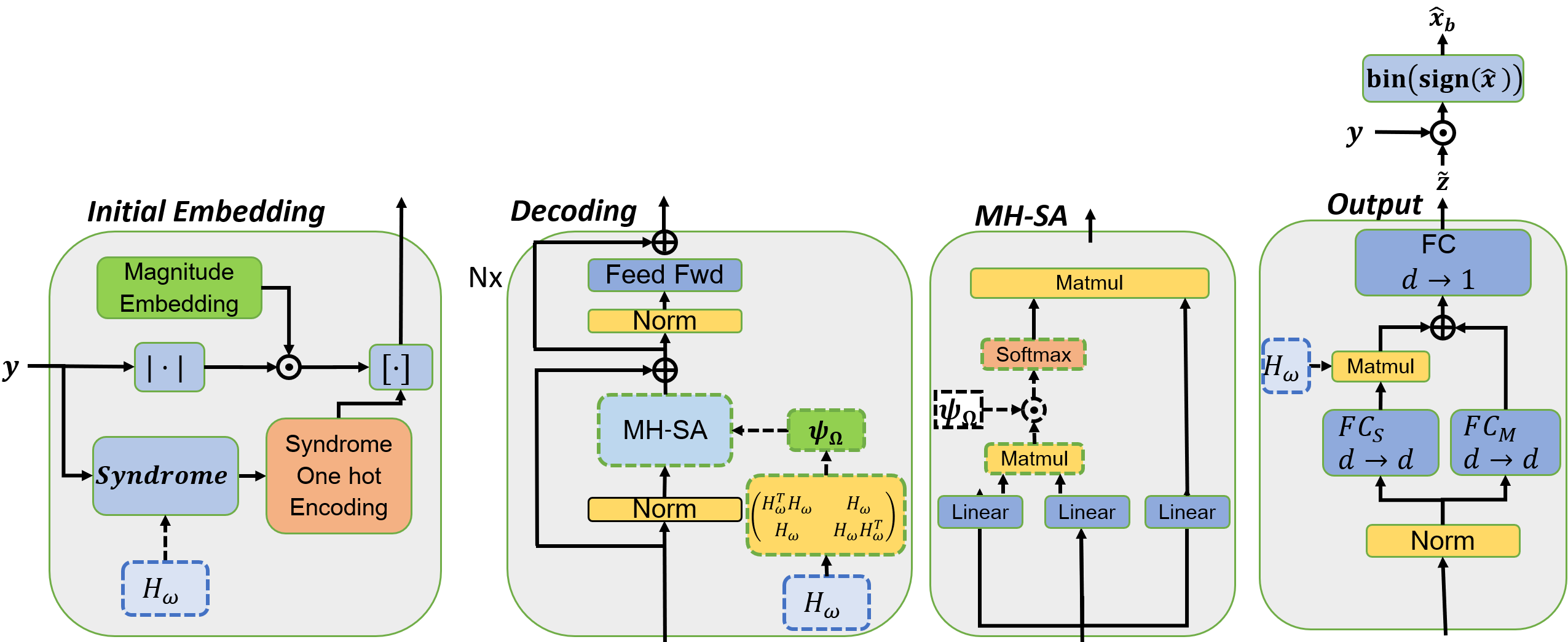}
\caption{Illustration of the proposed architecture. The main contributions are represented with dashed lines. 
}
\label{fig:e2e_ecct_arch}
\end{figure}

An illustration of the entire model is given in Figure \ref{fig:e2e_ecct_arch}. As with \cite{choukroun2024found}, the initial encoding is performed with a $d$ dimensional one-hot encoding for the syndrome part and a single $d$-dimensional vector for the magnitude part, for a total of three $d$-dimensional parameters.
Formally, the initial positional embedding $\Phi\in\mathbb{R}^{(2n-k)\times d}$ is given by 
\begin{equation}
\label{eq:ecct2}
{\Phi=[|y_{\Omega}|^{T} W_{m}},W_{s(y_{\Omega})}]
\end{equation}
with $W_{m}\in\mathbb{R}^d$ being the magnitude embedding vector and $W_{0}\in\mathbb{R}^d$ and $W_{1}\in\mathbb{R}^d$ the two one-hot encodings of each of the $n-k$ binary syndrome values.

The decoder is defined as a concatenation of $N$ decoding layers composed of self-attention and feed-forward layers interleaved with normalization layers with $d$. 
The distance embedding $\psi_{\gamma}:\mathbb{N}\rightarrow \mathbb{R}$ is a fully connected neural network with a 50-dimensional hidden layer and ReLU non-linearities mapping each number of paths to a scalar. This mapping becomes a fixed tensor at inference time. 
The contribution of each bit to itself (i.e. the diagonal elements) is omitted (masked) in the self-attention mechanism.

The output module is borrowed from \cite{choukroun2024found} to allow a code-aware prediction conditioned by the parameterized $H_{\Omega}$.
The output module performs the following projections on the final embedding $\Phi\coloneqq[\Phi_{M},\Phi_{S}]$
\begin{equation}
\begin{aligned}
\label{final_agg}
  \hat{\tilde{\varepsilon}} = \big(\Phi_{M}W_{M}+H_{\Omega}^{T} (\Phi_{S}W_{S})\big)W_{d\rightarrow 1}
\end{aligned}
\end{equation}
with $W_{S},W_{M}\in \mathbb{R}^{d\times d}$ and $W_{d\rightarrow 1}\in \mathbb{R}^{d}$ as the embedding layers.
Thus, our model remains code-, length-, and rate-invariant as well, by design.

The dimension of the feed-forward network of the transformer is four times that of the embedding $d$, following \cite{vaswani2017attention}, and is composed of GEGLU layers \citep{shazeer2020glu}, with layer normalization set to the pre-layer setting, as in \cite{opennmt,xiong2020layer}. 
An eight-head (i.e., $h=8$) self-attention module is used in all experiments. We note that while larger architectures 
would enable better performance, deepening the accuracy gap over other methods (e.g., GPT-3 \citep{brown2020gpt3} operates successfully on 2K inputs with a similar Transformer model, but with $N=96,d=12K$), ECC requires rather light and shallow models to be deployed on edge devices. 

The computational complexity as well as the number of parameters of the method are the same as of the FECCT~\cite{choukroun2024found}, since the code binary matrices ($G$ and $H$) as well as the distance embedding functions $\psi_{\gamma}$ are fixed after training.
The number of parameters is defined by $\mathcal{O}(N d^{2})$. In comparison, the ECCT is not length invariant and has $\mathcal{O}(N d^{2}+nd)$ parameters.

\begin{table*}[t]
    \centering
    \caption{
    A comparison of the negative natural logarithm of Bit Error Rate (BER) for three normalized SNR values of our method with literature baselines. 
	BP results are obtained after $L=5$ BP iterations in first row and \emph{at convergence} results in the second row are obtained after $L=50$ BP iterations. 
	SCL results are presented with a list length of $L=1$ in the first row and $L=32$ in the second row.
	Our performance is presented with fixed $\Omega$ (DC-ECCT) and with trained $\Omega$ (E2E DC-ECCT) for two shallow architectures: for $N=2,d=32$ in the first row and $N=6,d=128$ in the second row.
	The best results are in \textbf{bold}. The second best results are in \textit{italic}.
	Higher is better.
    }
    \label{tab:pretrained_table}
    \resizebox{0.65\textwidth}{!}{%
    \begin{tabular}{lc@{~}c@{~}cc@{~}c@{~}cc@{~}c@{~}cc@{~}c@{~}cc@{~}c@{~}c}
    \toprule
        Method & \multicolumn{3}{c}{BP} & \multicolumn{3}{c}{SCL} & \multicolumn{3}{c}{ECCT} & \multicolumn{3}{c}{DC-ECCT}  & \multicolumn{3}{c}{E2E DC-ECCT} \\
        \cmidrule(lr){2-4}
        \cmidrule(lr){5-7}
        \cmidrule(lr){8-10}
        \cmidrule(lr){11-13}
		\cmidrule(lr){14-16}
         $E_{b}/N_{0}$ & 4 & 5 & 6 & 4 & 5 & 6 & 4 & 5 & 6 & 4 & 5 & 6 & 4 & 5 & 6 \\ 
        \midrule                                                                                                                                                                                                                                                                                                                                                                                                                                      
		POLAR(32,11)	 		 	& \makecell{3.29\\ 3.84} 		& \makecell{3.77\\ 4.71}		& \makecell{4.21\\ 5.70}		
        \comment{NextMethod}		& \makecell{6.22\\ \textit{6.45}} 		& \makecell{8.06\\ \textit{8.37}}		& \makecell{10.28\\ \textit{10.60}}			
        \comment{NextMethod}		& \makecell{4.46\\ 6.37} 		& \makecell{5.57\\ 8.12}		& \makecell{7.01\\ 10.19}			
        \comment{NextMethod}		& \makecell{4.53\\ 6.41} 		& \makecell{5.69\\ 8.09}		& \makecell{7.08\\ 10.57}			
        \comment{NextMethod}		& \makecell{4.65\\ \textbf{6.62}} 		& \makecell{5.81\\ \textbf{8.57}}		& \makecell{7.28\\ \textbf{11.71}} \\
		\midrule                                                                                                                                                                                                                                                                                                                                                                                                                                                                                                   
		POLAR(64,32)	 		 	& \makecell{3.53\\ 4.29} 		& \makecell{4.02\\ 5.35}		& \makecell{4.45\\ 6.45}		
        \comment{NextMethod}		& \makecell{7.24\\ \textbf{8.16}} 		& \makecell{9.74\\ \textbf{10.73}}		& \makecell{12.91\\ \textbf{13.98}}			
        \comment{NextMethod}		& \makecell{4.56\\ 7.19} 		& \makecell{5.93\\ 9.70}		& \makecell{7.75\\ 13.33}			
        \comment{NextMethod}		& \makecell{4.40\\ 7.45} 		& \makecell{5.76\\ \textit{10.49}}		& \makecell{7.67\\ \textit{13.74}}			
        \comment{NextMethod}		& \makecell{4.72\\ \textit{7.59}} 		& \makecell{6.22\\ 10.38}		& \makecell{8.13\\ 13.09} \\
		\midrule                                                                                                    
		BCH(31,16)	 				& \makecell{4.59\\ 5.12} 		& \makecell{5.87\\ 6.87}		& \makecell{7.57\\ 9.27}		
        \comment{NextMethod}		& 		& NA		& 			
        \comment{NextMethod}		& \makecell{4.61\\ 6.37} 		& \makecell{5.97\\ 8.32}		& \makecell{7.69\\ 10.63}			
        \comment{NextMethod}		& \makecell{4.97\\ \textit{7.07}} 		& \makecell{6.56\\ \textbf{9.69}}		& \makecell{8.54\\ \textit{12.54}}			
        \comment{NextMethod}		& \makecell{5.30\\ \textbf{7.19}} 		& \makecell{6.89\\ \textit{{9.08}}}		& \makecell{9.09\\ \textbf{12.84}} \\
		\midrule                                                                                                                                                                                                                                                                                                                                                                                                                                                                                                   
		BCH(63,45)	 				& \makecell{4.07\\ 4.35} 		& \makecell{4.92\\ 5.60}		& \makecell{6.03\\ 7.24}		
        \comment{NextMethod}		& 		& NA		& 			
        \comment{NextMethod}		& \makecell{4.59\\ \textit{6.25}} 		& \makecell{6.07\\ \textit{8.78}}		& \makecell{8.13\\ \textit{12.45}}			
        \comment{NextMethod}		& \makecell{4.54\\ 6.08} 		& \makecell{6.07\\ 8.64}		& \makecell{8.14\\ 12.41}			
        \comment{NextMethod}		& \makecell{4.98\\ \textbf{6.37}} 		& \makecell{6.69\\ \textbf{9.09}}		& \makecell{8.97\\ \textbf{13.12}} \\
        \midrule                                                                                                                                                                                                                                                                                                                                                                                                                                                                                                   

		LDPC(49,24)	 			& \makecell{5.25\\ \textit{6.09}} 		& \makecell{7.15\\ \textit{8.75}}		& \makecell{9.86\\ \textit{11.91}}		
        \comment{NextMethod}		& 		& NA		& 			
        \comment{NextMethod}		& \makecell{4.21\\ 5.34} 		& \makecell{5.32\\ 6.43}		& \makecell{6.56\\ {7.21}}			
        \comment{NextMethod}		& \makecell{4.08\\ {5.57}} 		& \makecell{5.29\\ {6.55}}		& \makecell{6.55\\ { 7.21}}			
        \comment{NextMethod}		& \makecell{4.95\\ \textbf{6.28}} 		& \makecell{6.46\\ \textbf{8.77}}		& \makecell{8.33\\ \textbf{12.33}} \\

        \midrule                                                                                                                                                                                                                                                                                                        	
        RS(60,52)	 				& \makecell{4.43\\4.69} 		& \makecell{5.32\\ 6.43}		& \makecell{6.43\\ 7.56}		
        \comment{NextMethod}		& 		& NA		& 			
        \comment{NextMethod}		& \makecell{4.37\\ 4.37} 		& \makecell{5.11\\ 5.13}		& \makecell{6.03\\ 6.04}			
        \comment{NextMethod}		& \makecell{5.04\\ \textbf{5.61}} 		& \makecell{6.68\\ \textbf{7.59}}		& \makecell{8.82\\ \textit{9.82}}			
        \comment{NextMethod}		& \makecell{5.12\\ \textbf{5.61}} 		& \makecell{6.80\\ \textit{{7.56}}}		& \makecell{9.02\\ \textbf{9.90}} \\                                                                                                                                                                                           

		\bottomrule
	\end{tabular}
	}
\end{table*} 

\subsection{Training}
 The training objective is the cross-entropy function as given in Eq.~\ref{eq:e2e_obj}, while the estimated hard-decoded codeword is straightforwardly obtained as \mbox{$\hat{x}_{b} = \text{bin}(\text{sign}(f_{\theta}(h(y_{\Omega}))\cdot y_{\Omega})).$}

The Adam optimizer \citep{kingma2014adam} is used with 1024 samples per minibatch, for $1K$ epochs, with $1K$ minibatches per epoch. We initialized the learning rate to $10^{-4}$ coupled with a cosine decay scheduler down to $10^{-6}$ at the end of the training. No warmup was employed \citep{xiong2020layer}.
{\color{black}We note the optimization is stochastic and highly non-convex such that theoretical guarantees of the code performance or structure are difficult to establish.}

We observed that using a large batch size ($\times 8$) greatly improves the performance of our method, as well as of other baselines \cite{choukroun2022error,choukroun2024found}.
We, therefore, report in our tables new (and better) results for these baselines. We note that while using more epochs improves performance, the current modest setting already reaches state-of-the-art performance.

The initialization and optimization of $\Omega$ are of major importance in our highly non-convex optimization setting.
The initialization can be performed given a binary matrix $\Omega_{0}$ obtained from random sampling or from a baseline standardized parity-check matrix.
Thus, given an initial binary matrix $\Omega_{0}$, the learnable matrix is initialized as $\Omega=c\cdot\xi(\Omega_{0})$ with $c\in \mathbb{R}_{+}$ {, providing a uniform confidence to every element of $\Omega$.} 
{The hyperparameters used for the optimization of $\Omega$ are the early stopping training of $\Omega$ (not of the decoder), and the learning rate of $\Omega$ following \citep{courbariaux2015binaryconnect}. 
In all the experiments $c=0.01$, $\tau=\infty$ but with $\Omega$ clamped such that $|\Omega|<0.5$, as in \citep{courbariaux2015binaryconnect}. 
However, other values can be more optimal for any given code, and using larger batches would further improve performance.
}
As shown in Appendix \ref{appendix:rep_code}, it is important to train using the all ones message, i.e., $m=\mathbb{1}_{k}$.

Accelerating the proposed method (e.g. pruning, model quantization/binarization, distillation, low-rank approximation) \citep{wang2020linformer,lin2021survey} is beyond the scope of this paper and is left for future work. {\color{black}E.g., sparse self-attention can be induced via the regularization of the objective}. 
Training and experiments are performed on a 12GB GeForce RTX 2080 Ti GPU. The training time ranges from 34 to 327 seconds per epoch depending on code length, rate, and model size. Testing time ranges from 2 to 3ms per sample, using one GPU without any model optimization.

\begin{figure*}[t]
\centering
    \resizebox{0.885\textwidth}{!}
    {
\noindent  \begin{tabular}{@{}cccc@{}}
      (a) &(b) &(c) &(d) \\
\includegraphics[trim={0 0 0 0},clip, width=0.23\linewidth]{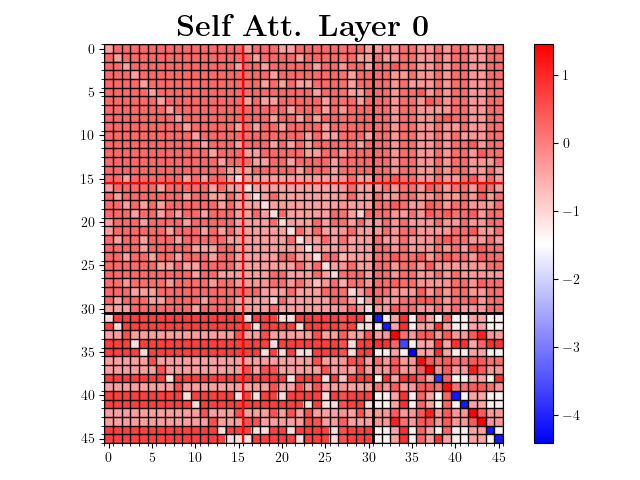} &
\includegraphics[trim={0 0 0 0},clip, width=0.23\linewidth]{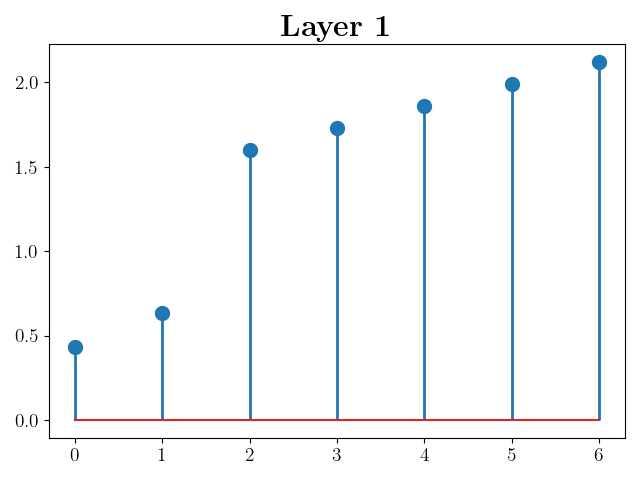} &
\includegraphics[trim={0 0 0 0},clip, width=0.23\linewidth]{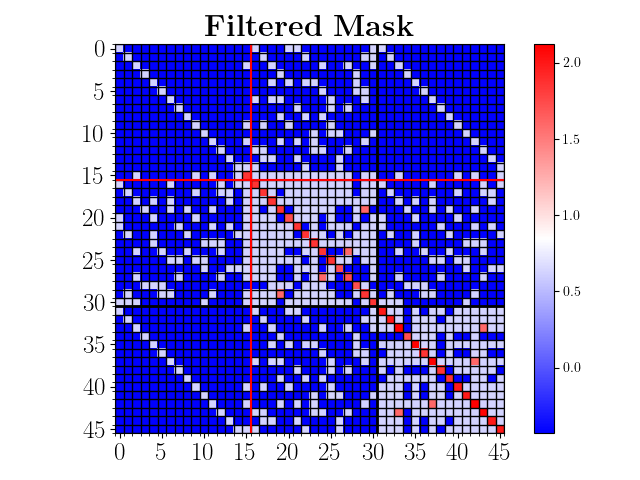} &
\includegraphics[trim={0 0 0 0},clip, width=0.23\linewidth]{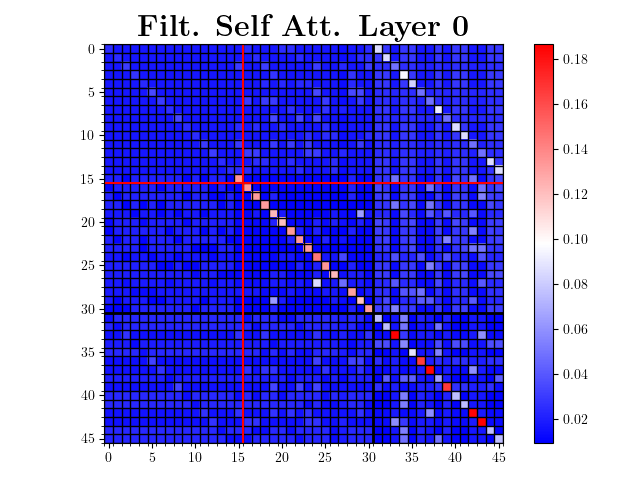} \\
\includegraphics[trim={0 0 0 0},clip, width=0.23\linewidth]{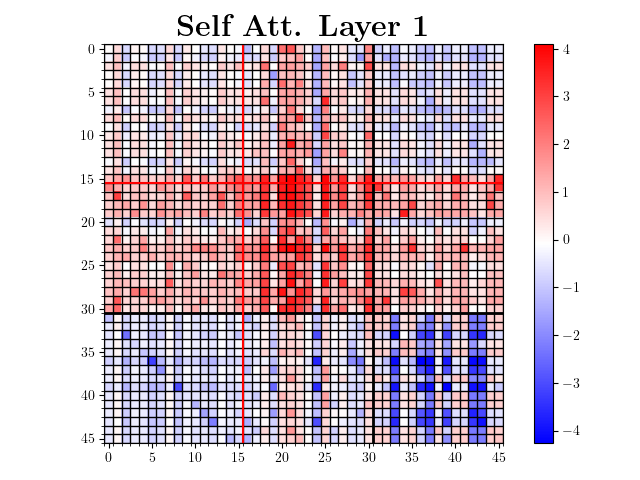} &
\includegraphics[trim={0 0 0 0},clip, width=0.23\linewidth]{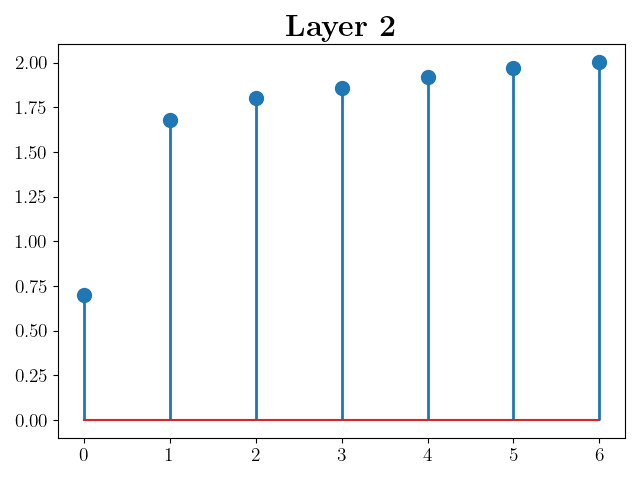} &
\includegraphics[trim={0 0 0 0},clip, width=0.23\linewidth]{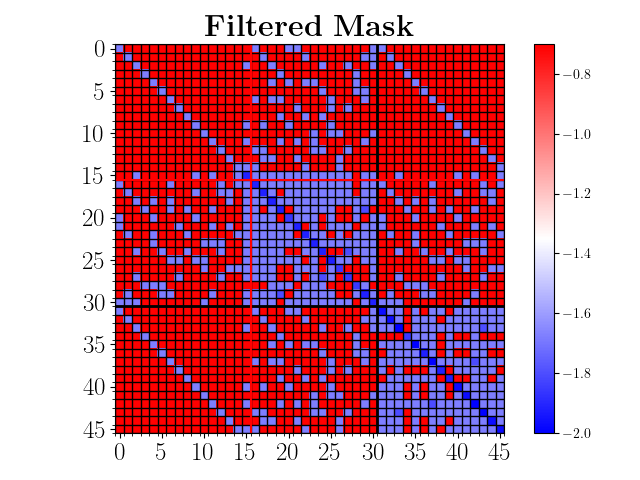} &
\includegraphics[trim={0 0 0 0},clip, width=0.23\linewidth]{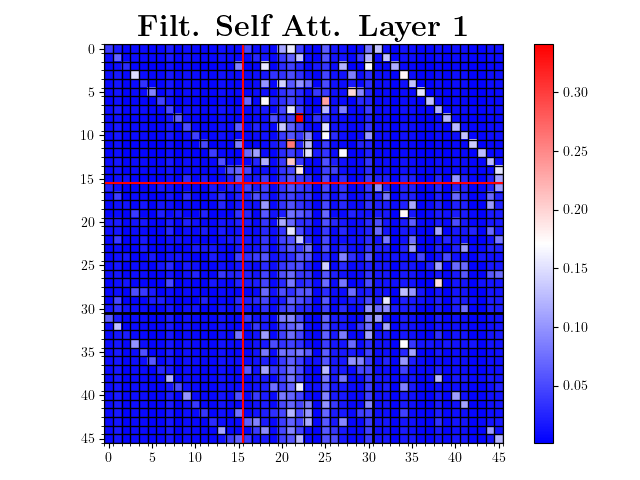} \\
      \end{tabular}
      }
  \caption{
  \textcolor{black}{
For a $N=2$ layers DC-ECCT (first and second row) and a (31,16) code: (a) self-attention layer, (b) connectivity mapping $\psi_{\gamma}$, (c) the corresponding filtered mask $\psi_{\gamma}\big(g(H_{\Omega})\big)$ (d) the obtained soft masked self-attention. The self-attention maps have been averaged over the heads dimension.}
}\label{fig:analysis_filters}
\end{figure*}
\vspace{-1em}

\section{Experiments}
To evaluate our method, we train the proposed architecture with four classes of linear codes: Low-Density Parity Check (LDPC) codes \citep{gallager1962low}, Polar codes \citep{arikan2008channel}, Reed Solomon codes \citep{reed1960polynomial} and Bose–Chaudhuri–Hocquenghem (BCH) codes \citep{bose1960class}.
All the parity check matrices are taken from \cite{channelcodes}. 
Code available at \url{https://github.com/yoniLc/E2E_DC_ECCT}.

We compare our method with the BP algorithm \citep{pearl1988probabilistic}, the SCL algorithm \cite{tal2015list} for polar codes, and the SOTA ECCT \citep{choukroun2022error} neural decoder. 
{FECCT is not tested, since while possessing important invariance properties, it reaches similar performance as the ECCT in average.  We note that, similarly to FECCT, our method can be implemented as a foundation model, where a single decoder is used for decoding multiple (potentially trained) codes. However, we are not looking to optimize multiple codes at once, due to the associated training complexity.} 
Other neural decoders are not presented, since their performance remains far from the ECCTs. 
Note that LDPC codes are specifically designed for BP-based decoding \cite{richardson2001design}.
SCL \cite{tal2015list} is specifically designed for (short) polar codes on which it is close to ML performance. 

As opposed to other methods \cite{nachmani2016learning,nachmani2019hyper,nachmani2021autoregressive}, our method as well as the other Transformer based neural decoders do not assume that the channel is known and do not then compute the LLRs as model input for improved performance. 
We observe that using LLR very slightly improves these methods, while removing it from classical BP induces catastrophic degradation of the performance.

The results are reported as negative natural logarithm bit error rates (BER), i.e., $-\ln(\text{BER})$, for three different normalized SNR values ($Eb/N_{0}$), following the conventional testing benchmark, e.g., \cite{nachmani2019hyper,choukroun2022error}. BP-based results are obtained after $L=5$ BP iterations in the first row (i.e., 10-layer neural network) and \emph{at convergence} results in the second row are obtained after $L=50$ BP iterations (i.e., 100-layer neural network). 
During testing, at least $10^{5}$ random codewords are decoded, to obtain at least $50$ frames with errors at each SNR value.  We trained and tested all reported ECCT results ourselves to ensure that the models were trained on the same parity-check matrices. {The SCL experiments are conducted by us, using the code framework of \citep{Cassagne2019a}.} 

{The hyperparameter search was performed using a validation set as follows. For all neural decoders (i.e., including ECCT), we selected the best results obtained from \emph{a single} random initialization and with the baseline initialization of $\Omega$. For the trained $\Omega$ setting, we tested the early stopping of $\Omega$ optimization after 800 epochs.
We also experimented with training the code with a smaller learning rate defined following \cite{courbariaux2015binaryconnect,glorot2010understanding}}.
Results showing the impact of the initialization on the proposed decoder and encoder-decoder framework are given in Appendix \ref{appendix:init}.
The parity-check matrices of all neural decoders are in standard form.


In Table \ref{tab:pretrained_table} we present the performance of our model Deep Coding Error Correction Code Transformer (DC-ECCT) on several codes. 
As can be seen, even for fixed (not trained) $\Omega$ our neural decoder outperforms the state-of-the-art neural decoder. Moreover, the end-to-end optimization of the code (E2E DC-ECCT) improves the performance by very large margins. We can observe that polar codes are already well-suited for the inductive bias of transformers. Also, on larger polar codes SCL gets close to ML and the code optimization seems to converge to a worse local minima.
{We provide BER plot visualizations in Appendix \ref{appendix:ber_vis}.}
\section{Ablation Study and Analysis}
We study the impact of the proposed method on other decoders and analyze the different modules of the method.
{ We provide in Appendix \ref{appendix:ablation} an ablation study of the different components of the proposed method, demonstrating the superiority of the components of the proposed solution.}

\subsection{Performance with Belief Propagation}
In Table \ref{tab:bp_res}, we present the performance of the Belief Propagation decoding algorithm on different codes and rates. We compare the performance of the baselines codes, these same codes in standard form, random codes, random standardized codes (i.e., random binary $\Omega$), and codes obtained from the $\Omega$ optimized via the E2E DC-ECCT.
We can observe that the learned codes outperform other codes under the BP decoder by very large margins, even if the code is presented in standard form.
This supports the claim that our method is able to provide good codes in a broader sense.

\begin{table}[h]
    \centering
    \caption{
    A comparison of the negative natural logarithm of BER for three normalized SNR values of the BP method on different codes {that also appeared in Table \ref{tab:pretrained_table}}. 
	Results are obtained after $L=5$ BP iterations in the first row and \emph{at convergence} results in the second row are obtained after $L=50$ BP iterations. Best for each $L$ in bold. 
    }
    \label{tab:bp_res}
    \resizebox{0.985\columnwidth}{!}
    {%
    \begin{tabular}{lc@{~}c@{~}cc@{~}c@{~}cc@{~}c@{~}cc@{~}c@{~}cc@{~}cc@{~}c@{~}c}
    \toprule
        Code & \multicolumn{3}{c}{Baseline} & \multicolumn{3}{c}{Standard form} & \multicolumn{3}{c}{Random $H$} & \multicolumn{3}{c}{Random $\Omega$} & \multicolumn{3}{c}{E2E $\Omega$}  \\
        \cmidrule(lr){2-4}
        \cmidrule(lr){5-7}
        \cmidrule(lr){8-10}
        \cmidrule(lr){11-13}
		\cmidrule(lr){14-16}
         $E_{b}/N_{0}$ & 4 & 5 & 6 & 4 & 5 & 6 & 4 & 5 & 6 & 4 & 5 & 6& 4 & 5 & 6 \\ 
        \midrule                                                                                                                                                                                                                                                                                                                                                                                                                                
		(31,16)	 		 	& \makecell{4.59\\ 5.12} 		& \makecell{5.87\\ 6.87}		& \makecell{7.57\\ 9.27}		
		\comment{NextMethod}		& \makecell{3.97\\ 4.64} 		& \makecell{4.67\\ 5.89}		& \makecell{5.57\\ 7.37}
        \comment{NextMethod}		& \makecell{3.23\\ 3.42} 		& \makecell{3.72\\ 4.18}		& \makecell{4.36\\ 5.33}			
        \comment{NextMethod}		& \makecell{4.22\\ 4.89} 		& \makecell{5.01\\ 6.14}		& \makecell{5.86\\ 7.38}			
        \comment{NextMethod}		& \textbf{\makecell{6.13\\ 6.42}} 		& \textbf{\makecell{7.95\\ 8.31}}		& \textbf{\makecell{9.90\\ 10.24}}	 \\
		\midrule                                                                                                    
    		(63,45)	 		 	     & \makecell{4.07\\ 4.35} 		& \makecell{4.92\\ 5.60}		& \makecell{6.03\\ 7.24}	
		\comment{NextMethod}		& \makecell{4.37\\ 4.78} 		& \makecell{5.33\\ 6.30}		& \makecell{6.42\\ 8.22}			
		\comment{NextMethod}		& \makecell{3.72\\ 3.93} 		& \makecell{4.26\\ 4.79}		& \makecell{4.93\\ 5.93}			
		\comment{NextMethod}		& \makecell{3.96\\ 4.27} 		& \makecell{4.61\\ 5.29}		& \makecell{5.41\\ 6.69}			
		\comment{NextMethod}		&\textbf{ \makecell{5.55\\ 6.15}} 		& \textbf{\makecell{7.33\\ 8.60}}		& \textbf{\makecell{9.23\\ 11.32}}	 \\
		\midrule                                                                                                                                                                                                                                                                                                                                                                                                                                                                                                   
          (60,52)	 		 	& \makecell{4.43\\ 4.69} 		& \makecell{5.32\\ 5.95}		& \makecell{6.43\\ 7.56}		
		  \comment{NextMethod}		& \makecell{4.60\\ 4.83} 		& \makecell{5.65\\ 6.21}		& \makecell{6.94\\ 8.00}
          \comment{NextMethod}		& \makecell{4.41\\ 4.65} 		& \makecell{5.32\\ 5.90}		& \makecell{6.45\\ 7.53}			
		  \comment{NextMethod}		& \makecell{4.60\\ 4.83} 		& \makecell{5.63\\ 6.19}		& \makecell{6.93\\ 7.97}			
		  \comment{NextMethod}		& \textbf{\makecell{4.73\\ 4.99}} 		& \textbf{\makecell{5.79\\ 6.45}}		& \textbf{\makecell{7.01\\ 8.35}}	 \\     
        \midrule 


		(64,32)	 		 	& \makecell{3.53\\ 4.29} 		& \makecell{4.02\\ 5.35}		& \makecell{4.45\\ 6.45}	
		\comment{NextMethod}		& \makecell{3.82\\ 4.81} 		& \makecell{4.37\\ 5.74}		& \makecell{5.03\\ 6.71}		
		\comment{NextMethod}		& \makecell{2.92\\ 2.96} 		& \makecell{3.37\\ 3.54}		& \makecell{3.73\\ 4.17}			
		\comment{NextMethod}		& \makecell{3.34\\ 3.59} 		& \makecell{3.90\\ 4.53}		& \makecell{4.64\\ 5.87}			
		\comment{NextMethod}		& \textbf{\makecell{6.93\\ 7.69}} 		& \textbf{\makecell{9.49\\ 10.04}}		& \textbf{\makecell{12.51\\ 12.94}}	 \\


		\bottomrule
	\end{tabular}
	}
\end{table}

In Table \ref{tab:ml_codes} we also present the ML decoding performance on the shortest codes, further demonstrating that the method can learn competitive codes, independently of the decoder.
\begin{table}[h]
    \centering
    \caption{
    A comparison of the maximum likelihood method on different codes.
	The code tested are the baseline codes, random code and, the proposed learned codes.
    }
    \label{tab:ml_codes}
    \resizebox{0.75\columnwidth}{!}
    {%
    \begin{tabular}{lc@{~}c@{~}cc@{~}c@{~}cc@{~}c@{~}c}
    \toprule
        Code & \multicolumn{3}{c}{Baseline} & \multicolumn{3}{c}{Random} & \multicolumn{3}{c}{E2E $\Omega$}  \\
        \cmidrule(lr){2-4}
        \cmidrule(lr){5-7}
        \cmidrule(lr){8-10}
         $E_{b}/N_{0}$ & 4 & 5 & 6 & 4 & 5 & 6 & 4 & 5 & 6 \\ 
        \midrule                                                                                                                                                                                                                                                                                                                                                                                                                                
		(31,16)	 		 			& 7.40 		& 9.81		& 13.11
		\comment{NextMethod}		& 7.25 		& 9.14		& 10.80
        \comment{NextMethod}		& 7.39 		& 9.54		& 12.19	 \\
		\midrule                                                                                                    
    	(32,11)	 		 	     	& 6.50 		& 8.28		& 10.71
		\comment{NextMethod}		& 7.08 		& 8.48		& 11.72
		\comment{NextMethod}		& 7.34 		& 9.48		& 11.79	 \\


		\bottomrule
	\end{tabular}
	}
\end{table}

\subsection{Parity-check Matrix Visualization}
\begin{figure}[t]
\centering
    \resizebox{0.95\columnwidth}{!}
    {
\noindent  \begin{tabular}{@{}cccc@{}}
\includegraphics[trim={0 0 0 0},clip, width=0.32\linewidth]{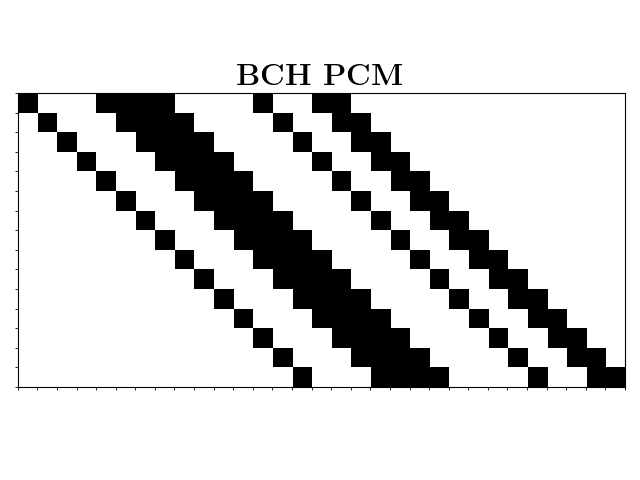} &
\includegraphics[trim={0 0 0 0},clip, width=0.32\linewidth]{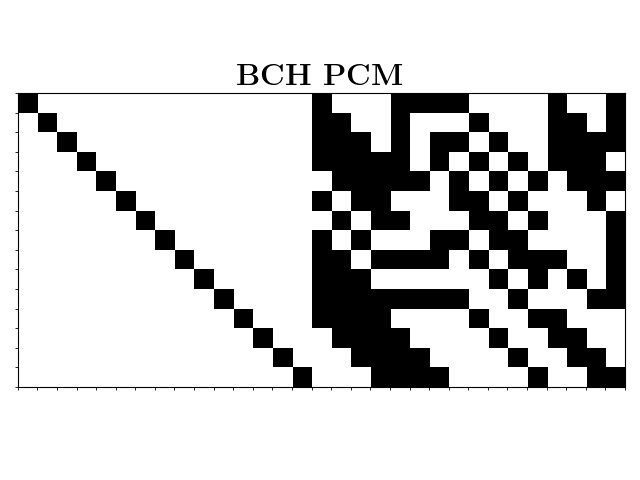} &
\includegraphics[trim={0 0 0 0},clip, width=0.32\linewidth] {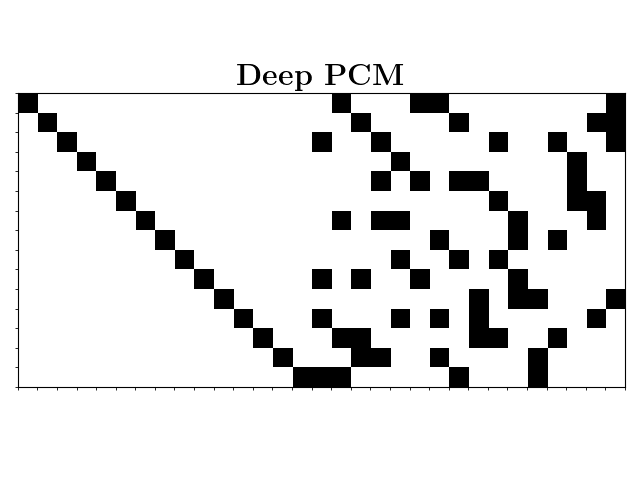} \\

      (a) &(b)&(c) \\
\includegraphics[trim={0 0 0 0},clip, width=0.32\linewidth]{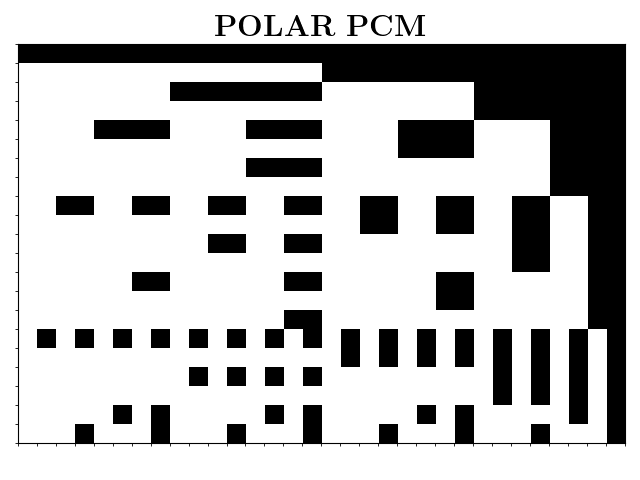} &
\includegraphics[trim={0 0 0 0},clip, width=0.32\linewidth]{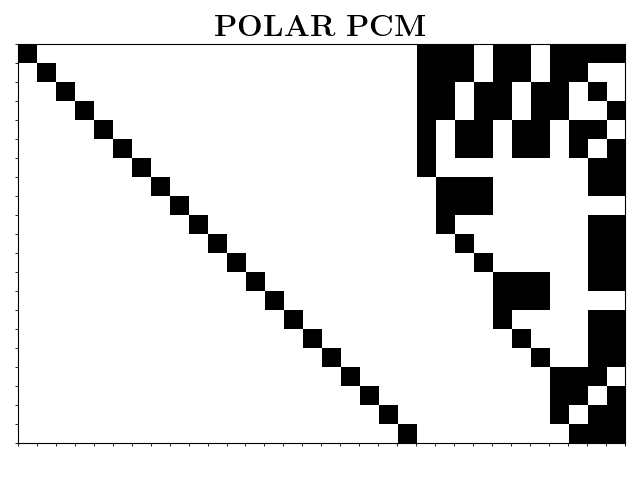} &
\includegraphics[trim={0 0 0 0},clip, width=0.32\linewidth]{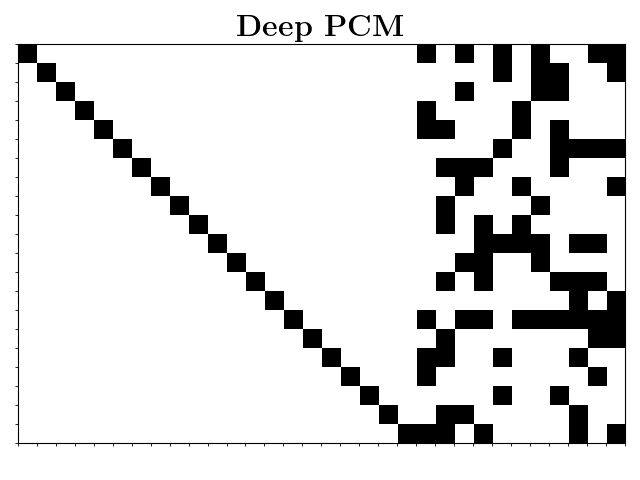} \\
(d) &(e)&(f) 
      \end{tabular}
      }
  \caption{
  \textcolor{black}{
The original parity-check matrix (PCM) of (a) BCH(31,16), (d) POLAR(32,11) and their standard form in (b) and (e), respectively. The third column corresponds to the learned parity-check matrices of the corresponding code length and rate.
The PCM sparsity is of (a) $25\%$, (b) $30\%$ (c) $16\%$, (d) $31\%$, (e) $17\%$, and (f) $15\%$.
}
}\label{fig:analysis_pc_mats}
\end{figure}

In Figure \ref{fig:analysis_pc_mats}, we depict several typical parity-check matrices for different codes, where we can see our optimization method {generally provides sparse codes.}
More visualizations {\color{black}and explanations} are given in Appendix \ref{appendix:pc_mat}.

\subsection{Visualization of Learned Mapping and Self-attention Maps}
In Figure \ref{fig:analysis_filters} we show the self-attention maps at the different layers of the model, at the different stages of the proposed soft masking. We first depict the classical self-attention layer in (a), then show the learned connectivity mapping $\psi_{\gamma}$ in (b), the induced filtered mask in (c), and the resulting masked/filtered self-attention (d).
We can observe from the connectivity mapping (b) that this shallow two-layer network learns to initially analyze highly connected nodes in the first layer, to finally focus on other less related nodes. This translates directly to complementary saliency regions of the filtered masks (c). Visualization for a $N=6$ model is given in Appendix 
\ref{appendix:sa_maps}.

\subsection{Training Dynamics}

\begin{figure}[t]
\centering
\noindent  \begin{tabular}{@{}cc@{}}
\includegraphics[trim={0 0 0 0},clip, width=0.48\linewidth]{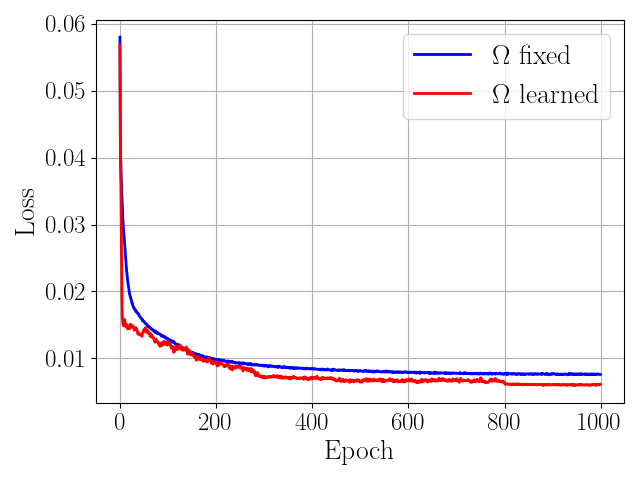} &
\includegraphics[trim={0 0 0 0},clip, width=0.48\linewidth]{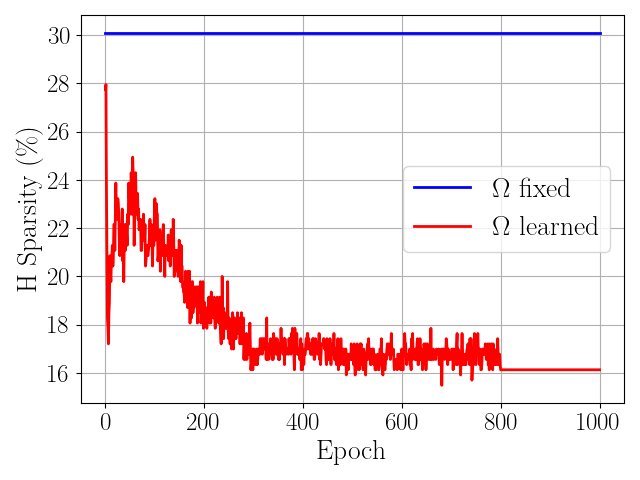} \\
     (a) &(b) \\
\includegraphics[trim={0 0 0 0},clip, width=0.48\linewidth]{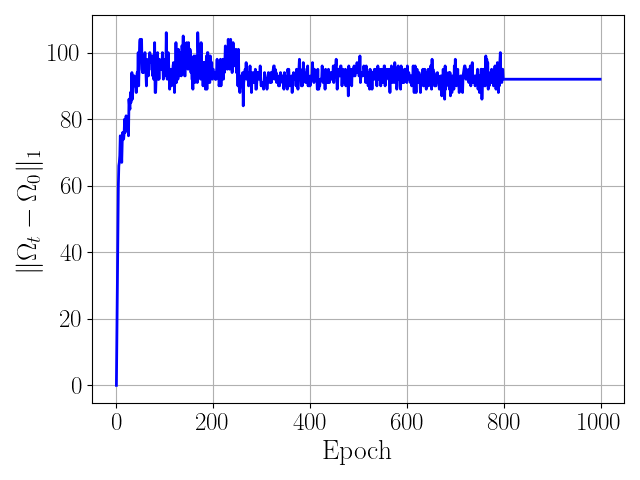} &
\includegraphics[trim={0 0 0 0},clip, width=0.48\linewidth]{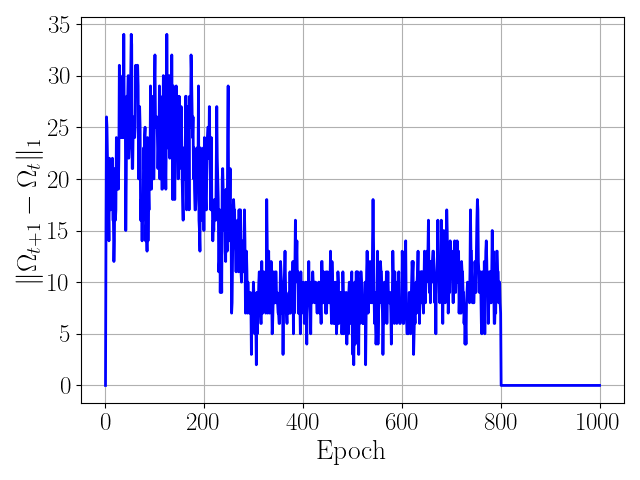} \\
(c) &(d) \\
\end{tabular}
  \caption{
  \textcolor{black}{ 
(a) Training loss of the proposed DC-ECCT with $\Omega$ fixed and trained for BCH(31,16) codes. (b) Evolution of $\Omega_{t}$ compared to $\Omega_{0}$. (c) Evolution of $\Omega_{t}$ compared to $\Omega_{t-1}$. (b) Parity-check matrix sparsity. In these experiments, the initial $\Omega_{0}$ is random,  the optimization over $\Omega$ is stopped at iteration 800, and the architecture is $N=2,d=32$.
}
}\label{fig:analysis_curves}
\end{figure}

We present in Figure \ref{fig:analysis_curves} the typical training dynamics of the proposed framework. In panel (a) we show the training loss for a fixed $\Omega$. As can be seen, the encoder-decoder models enable faster and better training.
We present in (b,c) the high variation of the learned $\Omega$ at the beginning of the optimization, attenuated towards the end of the training.
Finally, we can observe in (d) the level of sparsity of the code during training. The framework tends to produce sparse codes, with the main modification of the codes appearing during the first stage of training.

\section{Conclusion}
We present a novel end-two-end training method of the binary linear block error correction system. 
The proposed framework enables the effective and differentiable joint optimization of the code and of the neural decoder.
The neural decoder based on the Transformer architecture allows the differentiable training of the code via the Tanner graph connectivity derivation from the parity-check matrix.
The proposed neural decoder outperforms the state-of-the-art, while the unified encoding-decoding training allows further improvement of performance.

A common criticism for ML-based ECC is that the neural decoder cannot be deployed directly without the application of massive deep-learning acceleration methods. Here, we show for the first time that a code trained jointly with its decoder is also better for popular classical decoders. 
Looking forward, the efficient optimization of the codes may open the door to the creation of new families of codes and the establishment of new industry standards.

\newpage
\section{Impact Statements}
This paper presents work whose goal is to advance the field of Machine Learning and Information Theory. No real societal consequences of our work can be easily established or highlighted here.
\bibliography{main.bib}

\begin{thebibliography}{45}
\providecommand{\natexlab}[1]{#1}
\providecommand{\url}[1]{\texttt{#1}}
\expandafter\ifx\csname urlstyle\endcsname\relax
  \providecommand{\doi}[1]{doi: #1}\else
  \providecommand{\doi}{doi: \begingroup \urlstyle{rm}\Url}\fi

\bibitem[Aoudia \& Hoydis(2018)Aoudia and Hoydis]{aoudia2018end}
Aoudia, F.~A. and Hoydis, J.
\newblock End-to-end learning of communications systems without a channel model.
\newblock In \emph{2018 52nd Asilomar Conference on Signals, Systems, and Computers}, pp.\  298--303. IEEE, 2018.

\bibitem[Arikan(2008)]{arikan2008channel}
Arikan, E.
\newblock Channel polarization: A method for constructing capacity-achieving codes.
\newblock In \emph{2008 IEEE International Symposium on Information Theory}, pp.\  1173--1177. IEEE, 2008.

\bibitem[Bengio et~al.(2013)Bengio, L{\'e}onard, and Courville]{bengio2013estimating}
Bengio, Y., L{\'e}onard, N., and Courville, A.
\newblock Estimating or propagating gradients through stochastic neurons for conditional computation.
\newblock \emph{arXiv preprint arXiv:1308.3432}, 2013.

\bibitem[Bennatan et~al.(2018)Bennatan, Choukroun, and Kisilev]{bennatan2018deep}
Bennatan, A., Choukroun, Y., and Kisilev, P.
\newblock Deep learning for decoding of linear codes-a syndrome-based approach.
\newblock In \emph{2018 IEEE International Symposium on Information Theory (ISIT)}, pp.\  1595--1599. IEEE, 2018.

\bibitem[Bose \& Ray-Chaudhuri(1960)Bose and Ray-Chaudhuri]{bose1960class}
Bose, R.~C. and Ray-Chaudhuri, D.~K.
\newblock On a class of error correcting binary group codes.
\newblock \emph{Information and control}, 3\penalty0 (1):\penalty0 68--79, 1960.

\bibitem[Brown et~al.(2020)Brown, Mann, Ryder, Subbiah, Kaplan, Dhariwal, Neelakantan, Shyam, Sastry, Askell, et~al.]{brown2020gpt3}
Brown, T., Mann, B., Ryder, N., Subbiah, M., Kaplan, J.~D., Dhariwal, P., Neelakantan, A., Shyam, P., Sastry, G., Askell, A., et~al.
\newblock Language models are few-shot learners.
\newblock \emph{Advances in neural information processing systems}, 33:\penalty0 1877--1901, 2020.

\bibitem[Cammerer et~al.(2017)Cammerer, Gruber, Hoydis, and ten Brink]{cammerer2017scaling}
Cammerer, S., Gruber, T., Hoydis, J., and ten Brink, S.
\newblock Scaling deep learning-based decoding of polar codes via partitioning.
\newblock In \emph{GLOBECOM 2017-2017 IEEE Global Communications Conference}, pp.\  1--6. IEEE, 2017.

\bibitem[Cammerer et~al.(2022)Cammerer, Hoydis, Aoudia, and Keller]{cammerer2022graph}
Cammerer, S., Hoydis, J., Aoudia, F.~A., and Keller, A.
\newblock Graph neural networks for channel decoding.
\newblock In \emph{2022 IEEE Globecom Workshops (GC Wkshps)}, pp.\  486--491. IEEE, 2022.

\bibitem[Cassagne et~al.(2019)Cassagne, Hartmann, L\'eonardon, He, Leroux, Tajan, Aumage, Barthou, Tonnellier, Pignoly, {Le Gal}, and J\'ego]{Cassagne2019a}
Cassagne, A., Hartmann, O., L\'eonardon, M., He, K., Leroux, C., Tajan, R., Aumage, O., Barthou, D., Tonnellier, T., Pignoly, V., {Le Gal}, B., and J\'ego, C.
\newblock Aff3ct: A fast forward error correction toolbox!
\newblock \emph{Elsevier SoftwareX}, 10:\penalty0 100345, October 2019.
\newblock ISSN 2352-7110.
\newblock \doi{https://doi.org/10.1016/j.softx.2019.100345}.
\newblock URL \url{http://www.sciencedirect.com/science/article/pii/S2352711019300457}.

\bibitem[Choukroun \& Wolf(2022)Choukroun and Wolf]{choukroun2022error}
Choukroun, Y. and Wolf, L.
\newblock Error correction code transformer.
\newblock \emph{Advances in Neural Information Processing Systems (NeurIPS)}, 2022.

\bibitem[Choukroun \& Wolf(2023)Choukroun and Wolf]{choukroun2022zdenoising}
Choukroun, Y. and Wolf, L.
\newblock Denoising diffusion error correction codes.
\newblock In \emph{The Eleventh International Conference on Learning Representations (ICLR)}, 2023.

\bibitem[Choukroun \& Wolf(2024{\natexlab{a}})Choukroun and Wolf]{choukroun2023deep}
Choukroun, Y. and Wolf, L.
\newblock Deep quantum error correction.
\newblock In \emph{Proceedings of the AAAI Conference on Artificial Intelligence}, volume~38, pp.\  64--72, 2024{\natexlab{a}}.

\bibitem[Choukroun \& Wolf(2024{\natexlab{b}})Choukroun and Wolf]{choukroun2024found}
Choukroun, Y. and Wolf, L.
\newblock A foundation model for error correction codes.
\newblock In \emph{The Twelfth International Conference on Learning Representations (ICLR)}, 2024{\natexlab{b}}.
\newblock URL \url{https://openreview.net/forum?id=7KDuQPrAF3}.

\bibitem[Courbariaux et~al.(2015)Courbariaux, Bengio, and David]{courbariaux2015binaryconnect}
Courbariaux, M., Bengio, Y., and David, J.-P.
\newblock Binaryconnect: Training deep neural networks with binary weights during propagations.
\newblock \emph{Advances in neural information processing systems}, 28, 2015.

\bibitem[ESTI(2021)]{ETSI}
ESTI.
\newblock 5g nr multiplexing and channel coding. etsi 3gpp ts 38.212.
\newblock \url{https://www.etsi.org/deliver/etsi_ts/138200_138299/138212/16.02.00_60/ts_138212v160200p.pdf}, 2021.

\bibitem[Gallager(1962)]{gallager1962low}
Gallager, R.
\newblock Low-density parity-check codes.
\newblock \emph{IRE Transactions on information theory}, 8\penalty0 (1):\penalty0 21--28, 1962.

\bibitem[Glorot \& Bengio(2010)Glorot and Bengio]{glorot2010understanding}
Glorot, X. and Bengio, Y.
\newblock Understanding the difficulty of training deep feedforward neural networks.
\newblock In \emph{Proceedings of the thirteenth international conference on artificial intelligence and statistics}, pp.\  249--256. JMLR Workshop and Conference Proceedings, 2010.

\bibitem[Gruber et~al.(2017)Gruber, Cammerer, Hoydis, and ten Brink]{gruber2017deep}
Gruber, T., Cammerer, S., Hoydis, J., and ten Brink, S.
\newblock On deep learning-based channel decoding.
\newblock In \emph{2017 51st Annual Conference on Information Sciences and Systems (CISS)}, pp.\  1--6. IEEE, 2017.

\bibitem[Helmling et~al.(2019)Helmling, Scholl, Gensheimer, Dietz, Kraft, Ruzika, and Wehn]{channelcodes}
Helmling, M., Scholl, S., Gensheimer, F., Dietz, T., Kraft, K., Ruzika, S., and Wehn, N.
\newblock {D}atabase of {C}hannel {C}odes and {ML} {S}imulation {R}esults.
\newblock \url{www.uni-kl.de/channel-codes}, 2019.

\bibitem[Hoydis et~al.(2022)Hoydis, Cammerer, {Ait Aoudia}, Vem, Binder, Marcus, and Keller]{sionna}
Hoydis, J., Cammerer, S., {Ait Aoudia}, F., Vem, A., Binder, N., Marcus, G., and Keller, A.
\newblock Sionna: An open-source library for next-generation physical layer research.
\newblock \emph{arXiv preprint}, Mar. 2022.

\bibitem[Jiang et~al.(2019{\natexlab{a}})Jiang, Kannan, Kim, Oh, Asnani, and Viswanath]{jiang2019deepturbo}
Jiang, Y., Kannan, S., Kim, H., Oh, S., Asnani, H., and Viswanath, P.
\newblock Deepturbo: Deep turbo decoder.
\newblock In \emph{2019 IEEE 20th International Workshop on Signal Processing Advances in Wireless Communications (SPAWC)}, pp.\  1--5. IEEE, 2019{\natexlab{a}}.

\bibitem[Jiang et~al.(2019{\natexlab{b}})Jiang, Kim, Asnani, Kannan, Oh, and Viswanath]{jiang2019turbo}
Jiang, Y., Kim, H., Asnani, H., Kannan, S., Oh, S., and Viswanath, P.
\newblock Turbo autoencoder: Deep learning based channel codes for point-to-point communication channels.
\newblock \emph{Advances in neural information processing systems}, 32, 2019{\natexlab{b}}.

\bibitem[Kim et~al.(2018{\natexlab{a}})Kim, Jiang, Kannan, Oh, and Viswanath]{kim2018deepcode}
Kim, H., Jiang, Y., Kannan, S., Oh, S., and Viswanath, P.
\newblock Deepcode: Feedback codes via deep learning.
\newblock In \emph{Advances in Neural Information Processing Systems (NIPS)}, pp.\  9436--9446, 2018{\natexlab{a}}.

\bibitem[Kim et~al.(2018{\natexlab{b}})Kim, Jiang, Rana, Kannan, Oh, and Viswanath]{kim2018communication}
Kim, H., Jiang, Y., Rana, R., Kannan, S., Oh, S., and Viswanath, P.
\newblock Communication algorithms via deep learning.
\newblock In \emph{Sixth International Conference on Learning Representations (ICLR)}, 2018{\natexlab{b}}.

\bibitem[Kingma \& Ba(2014)Kingma and Ba]{kingma2014adam}
Kingma, D.~P. and Ba, J.
\newblock Adam: A method for stochastic optimization.
\newblock \emph{arXiv preprint arXiv:1412.6980}, 2014.

\bibitem[Klein et~al.(2017)Klein, Kim, Deng, Senellart, and Rush]{opennmt}
Klein, G., Kim, Y., Deng, Y., Senellart, J., and Rush, A.~M.
\newblock Opennmt: Open-source toolkit for neural machine translation.
\newblock In \emph{Proc. ACL}, 2017.
\newblock \doi{10.18653/v1/P17-4012}.
\newblock URL \url{https://doi.org/10.18653/v1/P17-4012}.

\bibitem[Kwak et~al.(2023)Kwak, Yun, Kim, Kim, and No]{kwak2023boosting}
Kwak, H.-Y., Yun, D.-Y., Kim, Y., Kim, S.-H., and No, J.-S.
\newblock Boosting learning for ldpc codes to improve the error-floor performance.
\newblock \emph{arXiv preprint arXiv:2310.07194}, 2023.

\bibitem[Lin et~al.(2021)Lin, Wang, Liu, and Qiu]{lin2021survey}
Lin, T., Wang, Y., Liu, X., and Qiu, X.
\newblock A survey of transformers.
\newblock \emph{arXiv preprint arXiv:2106.04554}, 2021.

\bibitem[Nachmani \& Wolf(2019)Nachmani and Wolf]{nachmani2019hyper}
Nachmani, E. and Wolf, L.
\newblock Hyper-graph-network decoders for block codes.
\newblock In \emph{Advances in Neural Information Processing Systems}, pp.\  2326--2336, 2019.

\bibitem[Nachmani \& Wolf(2021)Nachmani and Wolf]{nachmani2021autoregressive}
Nachmani, E. and Wolf, L.
\newblock Autoregressive belief propagation for decoding block codes.
\newblock \emph{arXiv preprint arXiv:2103.11780}, 2021.

\bibitem[Nachmani et~al.(2016)Nachmani, Be'ery, and Burshtein]{nachmani2016learning}
Nachmani, E., Be'ery, Y., and Burshtein, D.
\newblock Learning to decode linear codes using deep learning.
\newblock In \emph{2016 54th Annual Allerton Conference on Communication, Control, and Computing (Allerton)}, pp.\  341--346. IEEE, 2016.

\bibitem[O'Shea \& Hoydis(2017)O'Shea and Hoydis]{AutoencoderComm}
O'Shea, T.~J. and Hoydis, J.
\newblock An introduction to machine learning communications systems.
\newblock \emph{arXiv preprint arXiv:1702.00832}, 2017.

\bibitem[Park et~al.(2023)Park, Kwak, Kim, Kim, Kim, and No]{park2023mask}
Park, S.-J., Kwak, H.-Y., Kim, S.-H., Kim, S., Kim, Y., and No, J.-S.
\newblock How to mask in error correction code transformer: Systematic and double masking.
\newblock \emph{arXiv preprint arXiv:2308.08128}, 2023.

\bibitem[Pearl(1988)]{pearl1988probabilistic}
Pearl, J.
\newblock \emph{Probabilistic reasoning in intelligent systems: networks of plausible inference}.
\newblock Morgan kaufmann, 1988.

\bibitem[Raviv et~al.(2020)Raviv, Caciularu, Raviv, Goldberger, and Be'ery]{raviv2020graph}
Raviv, N., Caciularu, A., Raviv, T., Goldberger, J., and Be'ery, Y.
\newblock perm2vec: Graph permutation selection for decoding of error correction codes using self-attention.
\newblock \emph{arXiv preprint arXiv:2002.02315}, 2020.

\bibitem[Raviv et~al.(2023)Raviv, Goldmann, Vayner, Be'ery, and Shlezinger]{raviv2023crc}
Raviv, T., Goldmann, A., Vayner, O., Be'ery, Y., and Shlezinger, N.
\newblock Crc-aided learned ensembles of belief-propagation polar decoders.
\newblock \emph{arXiv preprint arXiv:2301.06060}, 2023.

\bibitem[Reed \& Solomon(1960)Reed and Solomon]{reed1960polynomial}
Reed, I.~S. and Solomon, G.
\newblock Polynomial codes over certain finite fields.
\newblock \emph{Journal of the society for industrial and applied mathematics}, 8\penalty0 (2):\penalty0 300--304, 1960.

\bibitem[Richardson et~al.(2001)Richardson, Shokrollahi, and Urbanke]{richardson2001design}
Richardson, T.~J., Shokrollahi, M.~A., and Urbanke, R.~L.
\newblock Design of capacity-approaching irregular low-density parity-check codes.
\newblock \emph{IEEE transactions on information theory}, 47\penalty0 (2):\penalty0 619--637, 2001.

\bibitem[Shannon(1948)]{shannon1948mathematical}
Shannon, C.~E.
\newblock A mathematical theory of communication.
\newblock \emph{The Bell system technical journal}, 27\penalty0 (3):\penalty0 379--423, 1948.

\bibitem[Shazeer(2020)]{shazeer2020glu}
Shazeer, N.
\newblock Glu variants improve transformer.
\newblock \emph{arXiv preprint arXiv:2002.05202}, 2020.

\bibitem[Tal \& Vardy(2015)Tal and Vardy]{tal2015list}
Tal, I. and Vardy, A.
\newblock List decoding of polar codes.
\newblock \emph{IEEE Transactions on Information Theory}, 61\penalty0 (5):\penalty0 2213--2226, 2015.

\bibitem[Vaswani et~al.(2017)Vaswani, Shazeer, Parmar, Uszkoreit, Jones, Gomez, Kaiser, and Polosukhin]{vaswani2017attention}
Vaswani, A., Shazeer, N., Parmar, N., Uszkoreit, J., Jones, L., Gomez, A.~N., Kaiser, {\L}., and Polosukhin, I.
\newblock Attention is all you need.
\newblock In \emph{Advances in neural information processing systems}, pp.\  5998--6008, 2017.

\bibitem[Wang et~al.(2020)Wang, Li, Khabsa, Fang, and Ma]{wang2020linformer}
Wang, S., Li, B.~Z., Khabsa, M., Fang, H., and Ma, H.
\newblock Linformer: Self-attention with linear complexity.
\newblock \emph{arXiv preprint arXiv:2006.04768}, 2020.

\bibitem[Xiong et~al.(2020)Xiong, Yang, He, Zheng, Zheng, Xing, Zhang, Lan, Wang, and Liu]{xiong2020layer}
Xiong, R., Yang, Y., He, D., Zheng, K., Zheng, S., Xing, C., Zhang, H., Lan, Y., Wang, L., and Liu, T.
\newblock On layer normalization in the transformer architecture.
\newblock In \emph{International Conference on Machine Learning}, pp.\  10524--10533. PMLR, 2020.

\bibitem[Ye et~al.(2018)Ye, Li, Juang, and Sivanesan]{ye2018channel}
Ye, H., Li, G.~Y., Juang, B.-H.~F., and Sivanesan, K.
\newblock Channel agnostic end-to-end learning based communication systems with conditional gan.
\newblock In \emph{2018 IEEE Globecom Workshops (GC Wkshps)}, pp.\  1--5. IEEE, 2018.

\end{thebibliography}
\bibliographystyle{icml2024}
\newpage
\appendix
\section{Repetition Code Backpropagation Analysis}
\label{appendix:rep_code}

We assume the standardized (3,1) repetition code such that 
$G=\begin{pmatrix}
1 & c_{1} & c_{2}
\end{pmatrix}$
and $H=\begin{pmatrix}
c_{1} & 1 & 0\\
c_{2} & 0 & 1
\end{pmatrix}$, where $c_{1},c_{2} \in \{0,1\}$. The repetition code is optimal for $c_1 =c_2 = 1$.

Given a message $m\in \{0,1\}$ we have $x=mG=\begin{pmatrix} m & c_{1}m & c_{2}m \end{pmatrix}$. Assuming no modulation, under a binary erasure channel we have $y=\begin{pmatrix} y_1 & y_2 & y_3 \end{pmatrix}=x\oplus n$ with $n\in \{0,1\}^{3}$ the noise vector.
The code syndrome is then of the form 
\begin{equation}
\label{eq:appendix_1}
\begin{aligned}
s = 
\begin{pmatrix}
s_0 \\
s_1
\end{pmatrix}
=Hy=
\begin{pmatrix}
c_1 y_1 \oplus y_{2}\\
c_2 y_1 \oplus y_{3}
\end{pmatrix}
\end{aligned}
\end{equation}

Using polarized notation, we have 
\begin{equation}
\label{eq:appendix_2}
\begin{aligned}
\begin{cases}
s_1 &= 0.5-0.5(1-2c_1 y_1 )(1-2y_2 )\\
s_2 &= 0.5-0.5(1-2c_2 y_1 )(1-2y_3 )
\end{cases}
\end{aligned}
\end{equation}

Thus, we have $\frac{\partial s_{1}}{\partial c_{2}}=\frac{\partial s_{2}}{\partial c_{1}}=0$ and, without loss of generality we have
\begin{equation}
\label{eq:appendix_3}
\begin{aligned}
\frac{\partial s_{1}}{\partial c_{1}}=y_1 (1-2y_2 ) + (1-2c_1 y_1 )\frac{\partial y_{2}}{\partial c_{1}}
\end{aligned}
\end{equation}

We can observe that assuming $y_{i}$ as independent from the codeword (i.e., not derived from $G$) for \emph{every} (potentially wrong) value of $c_{i}$, the gradient is zero and no update is possible half of the time (i.e., $m=n_{1}=0$ or $m=n_{1}=1$). Thus the channel output should not be assumed as constant during backpropagation (i.e., $y_{i}\coloneqq y_{i}(c_{1},c_{2})$).

Assuming the channel output is dependent on the generator and considering it for the computation of the codeword we have now $y_{i}=x_{i}\oplus n_{i}$ and then
\begin{equation}
\label{eq:appendix_4}
\begin{aligned}
\frac{\partial s_{1}}{\partial c_{1}}=y_1 (1-2y_2 ) + m(1-2c_1 y_1 )(1-2 n_2 )
\end{aligned}
\end{equation}
We can observe sampling zero messages induces the same gradient as the independent setting since it cancels the parity check coupled with it. Thus, we can deduct the all ones \emph{binary} message i.e., $m=1$ (or at least $m\neq 0$, a contrary to what is commonly done in neural decoders training) is important for efficient backpropagation since it will not cancel the information propagation in the backward pass.

\section{Parity-check Matrix Visualization}
\label{appendix:pc_mat}
\begin{figure}[h]
\centering
    \resizebox{0.95\columnwidth}{!}
    {
\noindent  \begin{tabular}{@{}cc@{}}
\\
\includegraphics[trim={0 0 0 0},clip, width=0.48\linewidth]{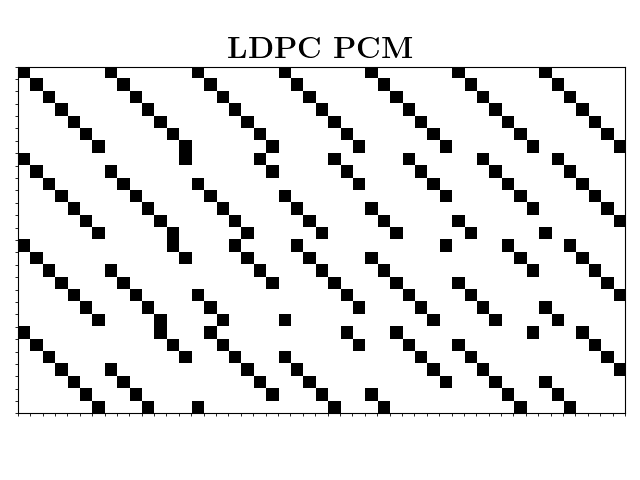} &
\includegraphics[trim={0 0 0 0},clip, width=0.48\linewidth]{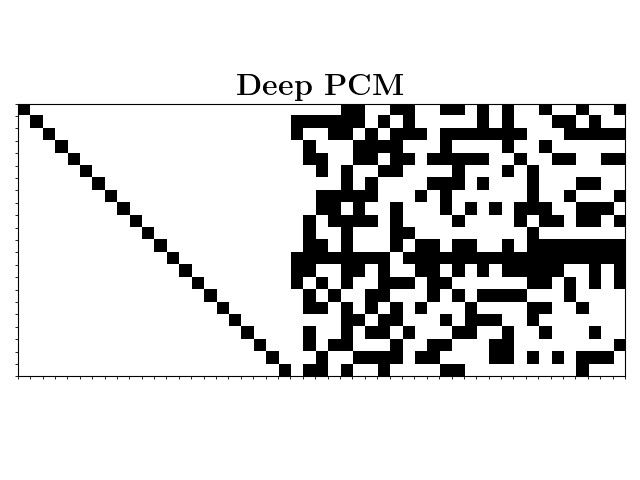} \\
(a) &(b) \\
\includegraphics[trim={0 0 0 0},clip, width=0.48\linewidth]{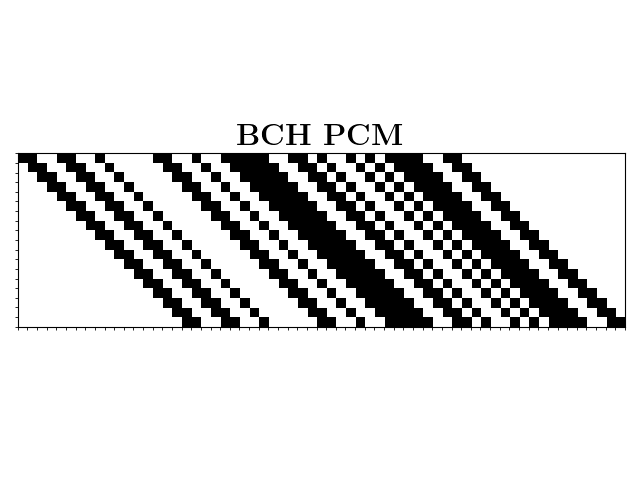} &
\includegraphics[trim={0 0 0 0},clip, width=0.48\linewidth]{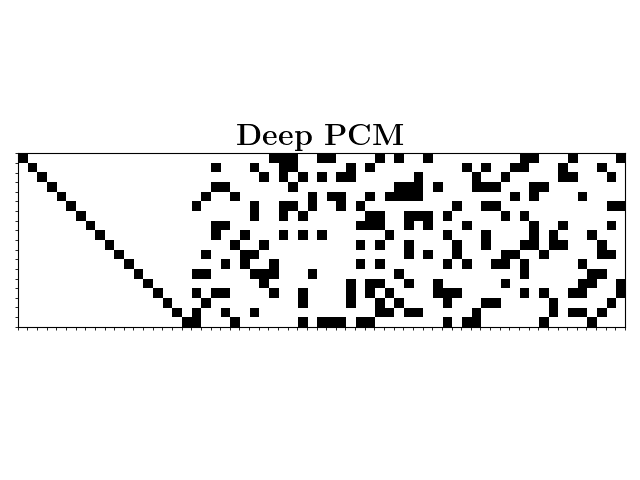} \\
(c) &(d) 
      \end{tabular}
      }
  \caption{
  \textcolor{black}{
The original parity-check matrix (PCM) of additional codes. (a) LDPC(49,24), and (c) BCH(63,45). The second column (b,d) corresponds to the learned parity-check matrices of the corresponding code length and rate.
The PCM sparsity is of (a) $15\%$, (b) $26\%$, (c) $38\%$, and (d) $20\%$ 
}
}\label{fig:app_analysis_pc_mats}
\end{figure}

In Figure \ref{fig:app_analysis_pc_mats}, we depict several typical parity-check matrices for different codes. 
{\color{black}We note that the sparsity of the code  is a property that emerges from training and is not enforced by the framework: sparse codes seem more appropriate for Transformer-based neural decoders' inductive bias.}

\section{Impact of the Initialization}
\label{appendix:init}
We present in Table \ref{tab:app_init} the performance of the DC-ECCT and the E2E DC-ECCT under baseline and random initialization. We can observe that the initialization has a very major impact on the DC-ECCT while the end-to-end training allows to mitigate the code impact but still presents different performance under different initialization, which is understandable in high dimensional non-convex optimization scenarios.
Also, it seems Polar codes, contrary to other codes, already provide a strong initialization for the proposed framework.

{\color{black}
 From our experience, refining a learned parity check matrix most of the time did not improve or even gave worse performance. The reason should lie within the high dimensional non-convex optimization where the local minima of 
 may be a sharp minimum [1] such that small perturbation (retraining) induces a worse local optimum. Other optimization methods may allow such refinement though.
}

\begin{table}[h]
    \centering
    \caption{
    A comparison of the negative natural logarithm of Bit Error Rate (BER) for three normalized SNR values of our method for $N=2,d=32$ with different initializations.
    We test the initialization of $\Omega$ with baseline codes and with random parity-check matrix. 
	Higher is better.
    }
    \label{tab:app_init}
    \resizebox{0.985\columnwidth}{!}{
    \begin{tabular}{lc@{~}c@{~}cc@{~}c@{~}cc@{~}c@{~}cc@{~}c@{~}c}
    \toprule
        Model & \multicolumn{6}{c}{DC-ECCT} & \multicolumn{6}{c}{E2E DC-ECCT}\\
        Code & \multicolumn{3}{c}{Baseline $\Omega$} & \multicolumn{3}{c}{Random $\Omega$} & \multicolumn{3}{c}{Baseline $\Omega$} & \multicolumn{3}{c}{Random $\Omega$}  \\
        \cmidrule(lr){2-4}
        \cmidrule(lr){5-7}
        \cmidrule(lr){8-10}
        \cmidrule(lr){11-13}
         $E_{b}/N_{0}$ & 4 & 5 & 6 & 4 & 5 & 6 & 4 & 5 & 6 & 4 & 5 & 6  \\ 
        \midrule                                                                                                                                                                                                                                                                                                                                                                                                                                      
		POLAR(32,11)	 		 	
            \comment{NextMethod}		& {4.53} 		& {5.69}		& {7.08} 
            \comment{NextMethod}		& {3.95} 		& {4.93}		& {6.18}
            \comment{NextMethod}		& {4.55} 		& {5.68}		& {7.08} 
            \comment{NextMethod}		& {4.65} 		& {5.81}		& {7.28}
        \\
		\midrule                                                                                                                                                                                                                                                                                                                                                                                                                                                                                                   
		POLAR(64,32)	 		 	
              \comment{NextMethod}		& {4.40} 		& {5.76}		& {7.57} 
            \comment{NextMethod}		& {3.47} 		& {4.42}		& {5.77}
            \comment{NextMethod}		& {4.72} 		& {6.22}		& {8.13} 
            \comment{NextMethod}		& {4.70} 		& {6.13}		& {8.05}
            \\
		\midrule                                                                                                    
		BCH(31,16)
              \comment{NextMethod}		& {4.97} 		& {6.56}		& {8.54} 
            \comment{NextMethod}		& {4.62} 		& {5.90}		& {7.59}
            \comment{NextMethod}		& {5.22} 		& {6.75}		& {8.56} 
            \comment{NextMethod}		& {5.30} 		& {6.89}		& {9.09}      \\
		\midrule                                                                                                                                                                                                                                                                                                                                                                                                                                                                                                   
		BCH(63,45)	 				
              \comment{NextMethod}		& {4.54} 		& {6.06}		& {8.12} 
            \comment{NextMethod}		& {4.41} 		& {5.85}		& {7.83}
            \comment{NextMethod}		& {4.72} 		& {6.37}		& {8.60} 
            \comment{NextMethod}		& {4.98} 		& {6.69}		& {8.97}      \\
        \midrule                                                                                                                                                                                                                                                                                                                                                                                                                                                                                                   

		LDPC(49,24)	 			
              \comment{NextMethod}		& {4.08} 		& {5.19}		& {6.45} 
            \comment{NextMethod}		& {3.75} 		& {4.81}		& {6.24}
            \comment{NextMethod}		& {4.95} 		& {6.45}		& {8.33} 
            \comment{NextMethod}		& {4.95} 		& {6.46}		& {8.33}      \\

        \midrule                                                                                                                                                                                                                                                                                                        	
        RS(60,52)	 				
                    \comment{NextMethod}		& {4.38} 		& {5.13}		& {6.03} 
            \comment{NextMethod}		& {5.04} 		& {6.68}		& {8.82}
            \comment{NextMethod}		& {4.38} 		& {5.13}		& {6.03} 
            \comment{NextMethod}		& {5.12} 		& {6.80}		& {9.02}\\                                                                                                                                                                                           

		\bottomrule
	\end{tabular}
	}
\end{table} 

\section{Visualization of Learned Mapping and Self-attention Maps}
\label{appendix:sa_maps}
\begin{figure*}[t]
\centering
    \resizebox{0.885\textwidth}{!}
    {
\noindent  \begin{tabular}{@{}cccc@{}}
(a) &(b) &(c) &(d) \\

\includegraphics[trim={0 0 0 0},clip, width=0.23\linewidth]{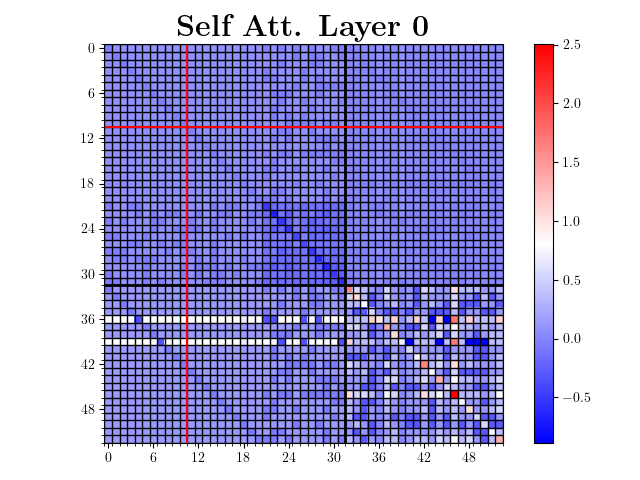}&
\includegraphics[trim={0 0 0 0},clip, width=0.23\linewidth]{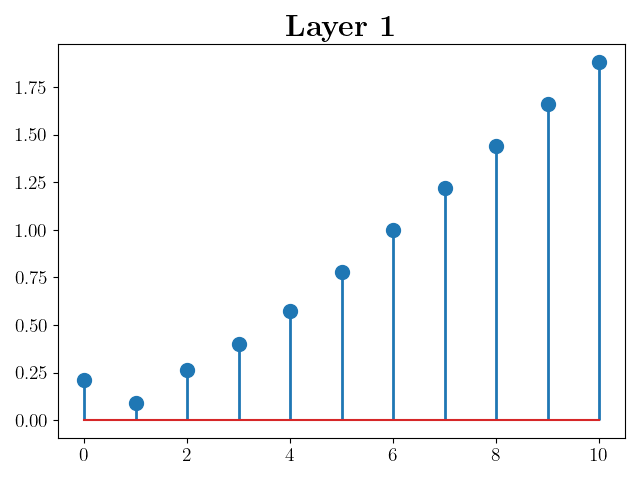} &
\includegraphics[trim={0 0 0 0},clip, width=0.23\linewidth]{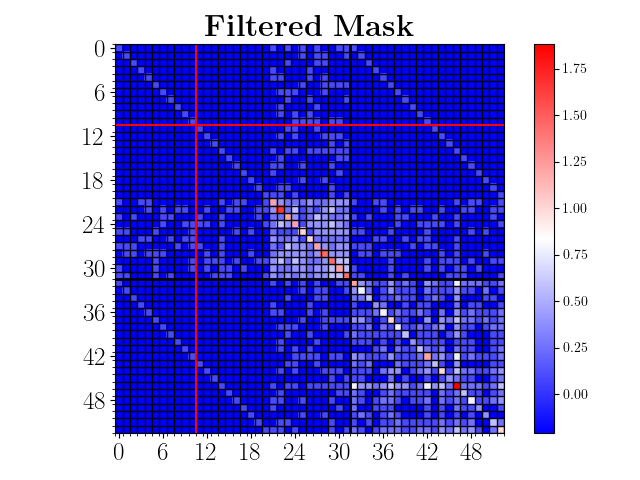} &
\includegraphics[trim={0 0 0 0},clip, width=0.23\linewidth]{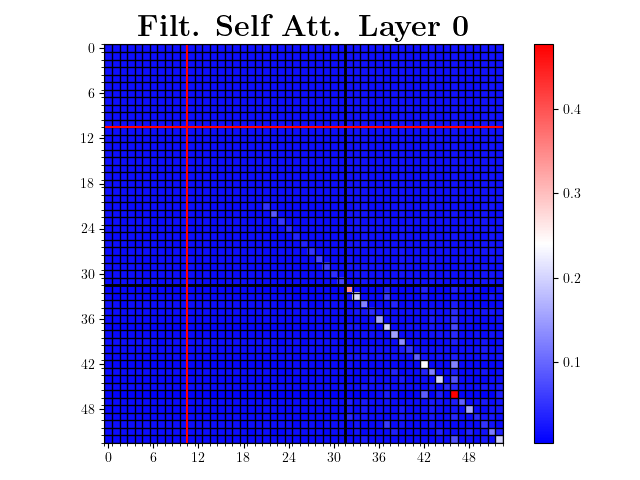} \\
\includegraphics[trim={0 0 0 0},clip, width=0.23\linewidth]{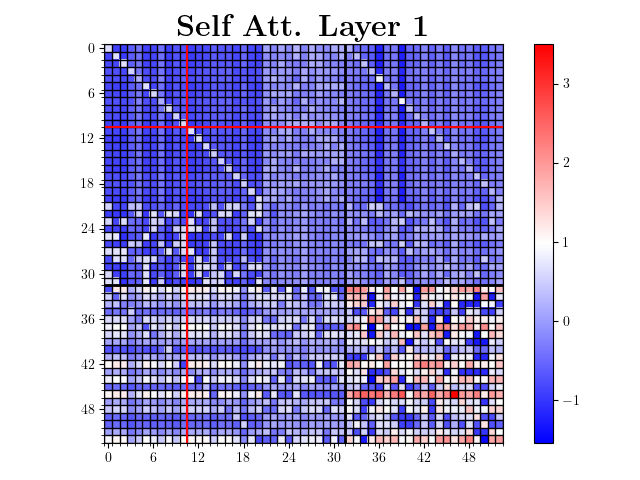}&
\includegraphics[trim={0 0 0 0},clip, width=0.23\linewidth]{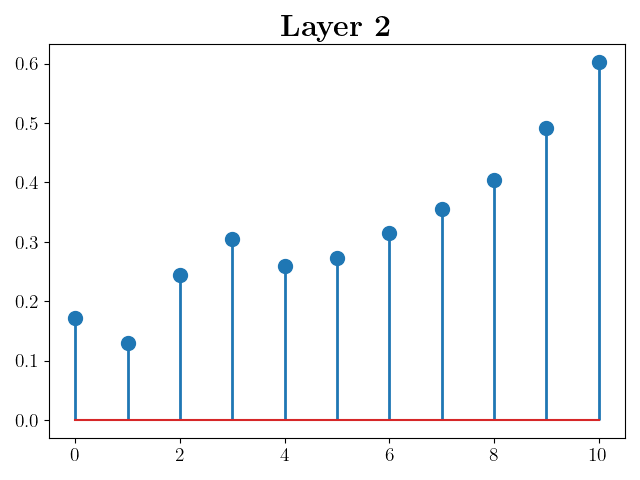} &
\includegraphics[trim={0 0 0 0},clip, width=0.23\linewidth]{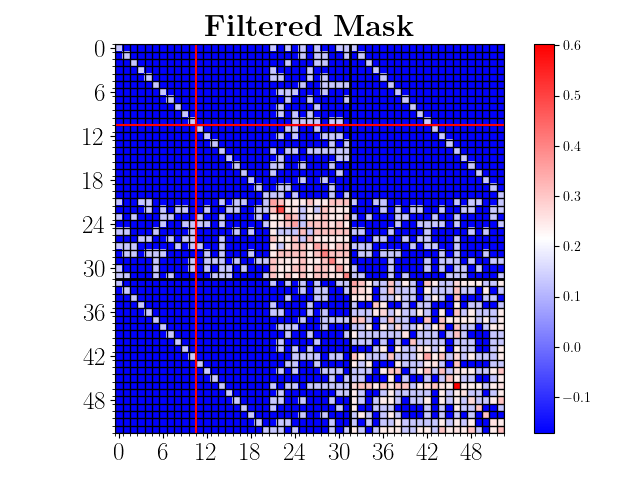} &
\includegraphics[trim={0 0 0 0},clip, width=0.23\linewidth]{appendix/sa_maps/sa_ebno_2_layer_filtered_0.png} \\
\includegraphics[trim={0 0 0 0},clip, width=0.23\linewidth]{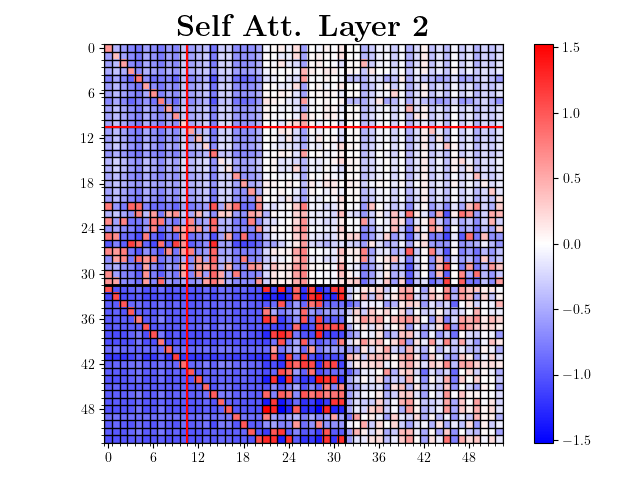}&
\includegraphics[trim={0 0 0 0},clip, width=0.23\linewidth]{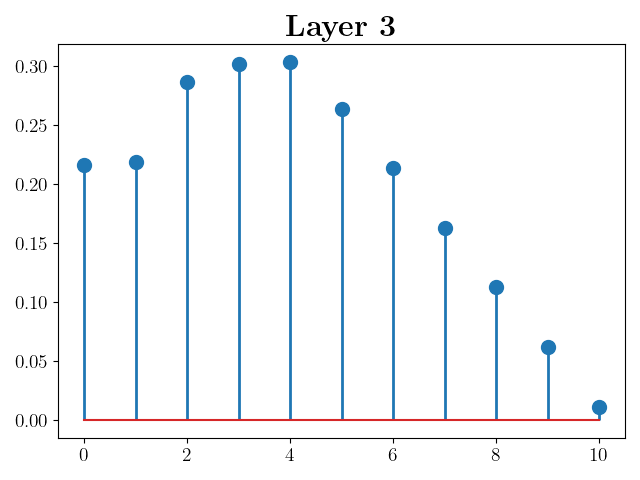} &
\includegraphics[trim={0 0 0 0},clip, width=0.23\linewidth]{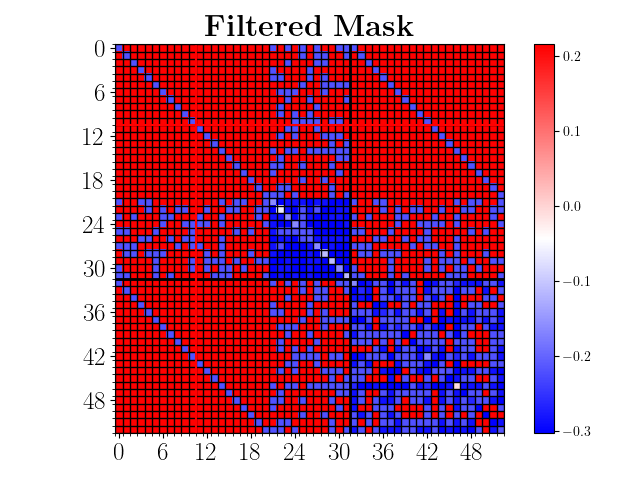} &
\includegraphics[trim={0 0 0 0},clip, width=0.23\linewidth]{appendix/sa_maps/sa_ebno_2_layer_filtered_0.png} \\
\includegraphics[trim={0 0 0 0},clip, width=0.23\linewidth]{appendix/sa_maps/sa_ebno_4_layer_2.png}&
\includegraphics[trim={0 0 0 0},clip, width=0.23\linewidth]{appendix/sa_maps/Deep_n_32_k_11_Layer_2.png} &
\includegraphics[trim={0 0 0 0},clip, width=0.23\linewidth]{appendix/sa_maps/Deep_n_32_k_11_FILTERED_MASK_Layer_0.png} &
\includegraphics[trim={0 0 0 0},clip, width=0.23\linewidth]{appendix/sa_maps/sa_ebno_2_layer_filtered_0.png} \\
\includegraphics[trim={0 0 0 0},clip, width=0.23\linewidth]{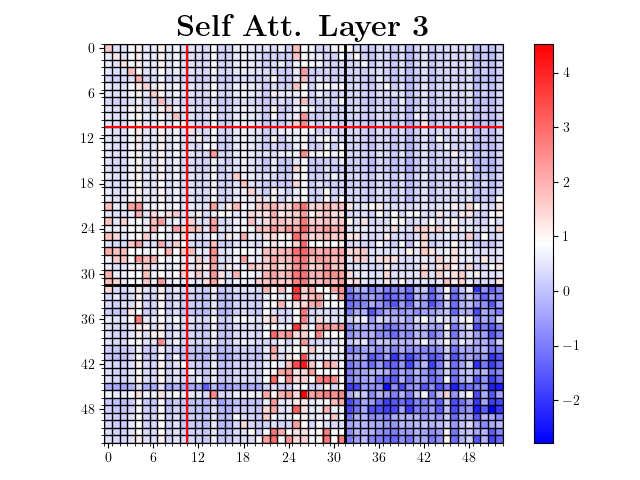}&
\includegraphics[trim={0 0 0 0},clip, width=0.23\linewidth]{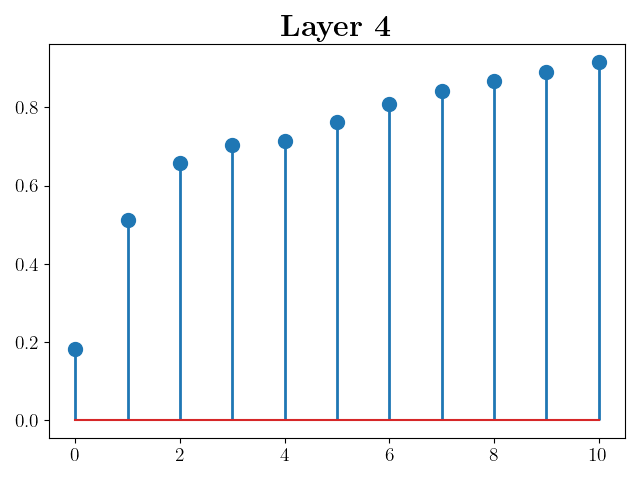} &
\includegraphics[trim={0 0 0 0},clip, width=0.23\linewidth]{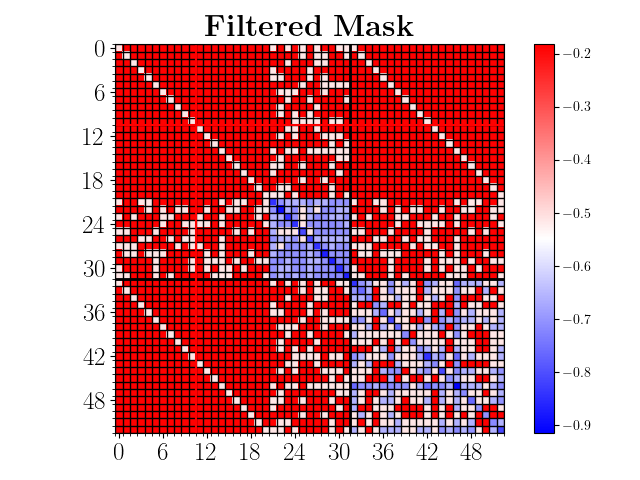} &
\includegraphics[trim={0 0 0 0},clip, width=0.23\linewidth]{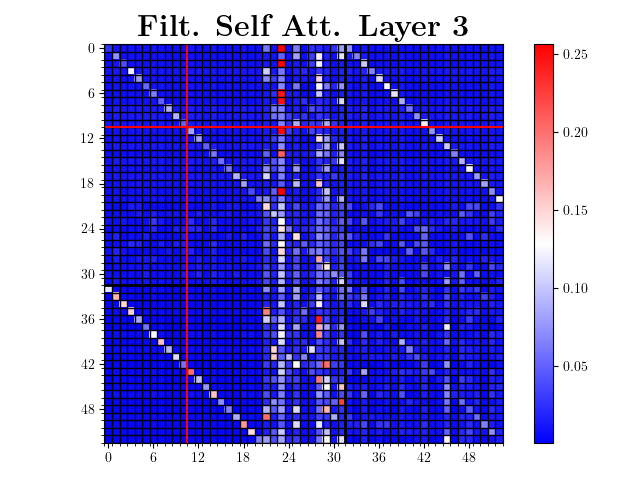} \\
\includegraphics[trim={0 0 0 0},clip, width=0.23\linewidth]{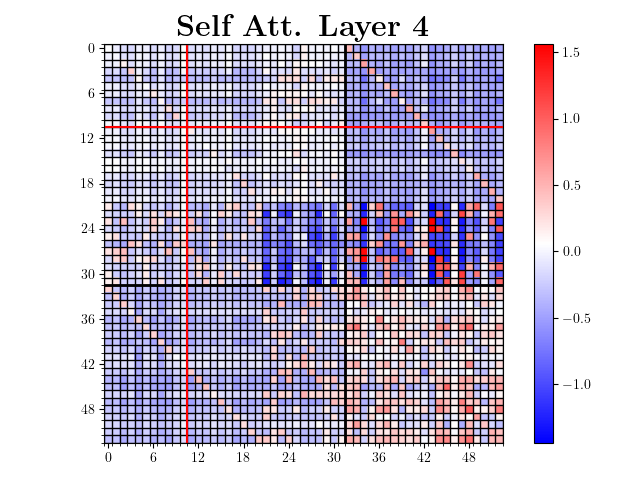}&
\includegraphics[trim={0 0 0 0},clip, width=0.23\linewidth]{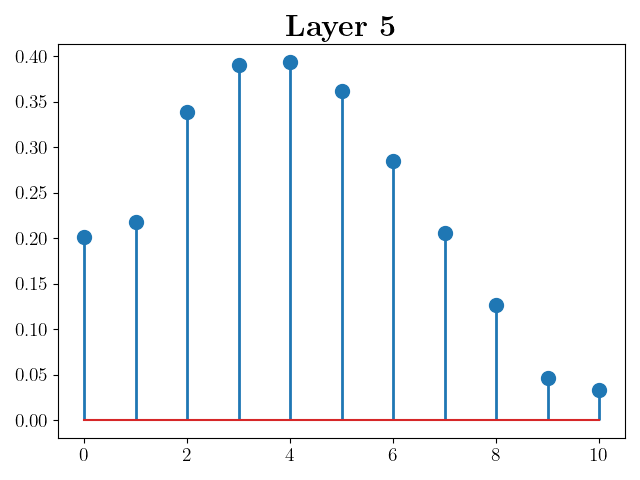} &
\includegraphics[trim={0 0 0 0},clip, width=0.23\linewidth]{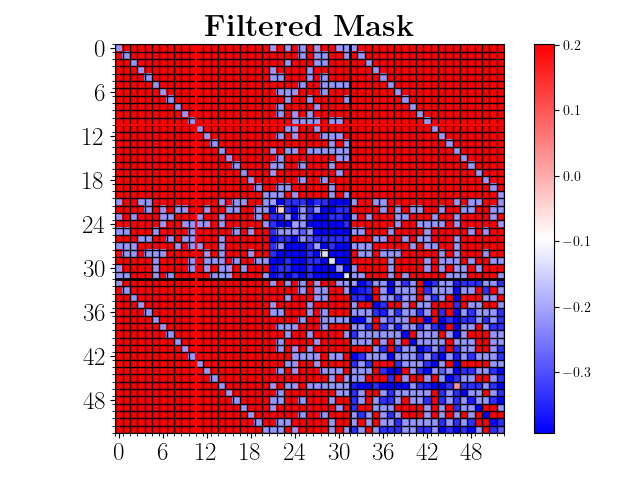} &
\includegraphics[trim={0 0 0 0},clip, width=0.23\linewidth]{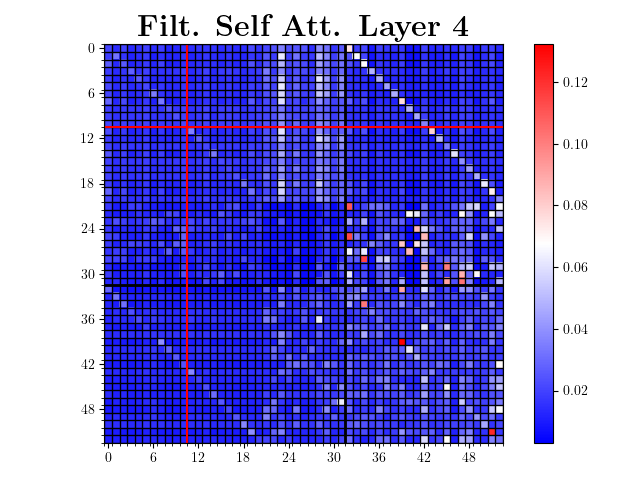} \\
\includegraphics[trim={0 0 0 0},clip, width=0.23\linewidth]{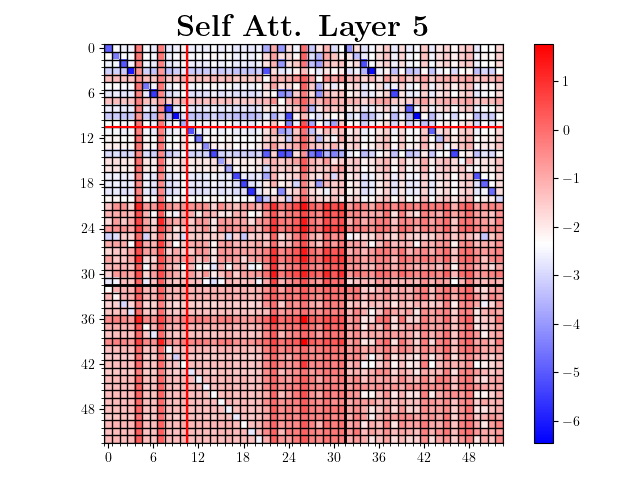}&
\includegraphics[trim={0 0 0 0},clip, width=0.23\linewidth]{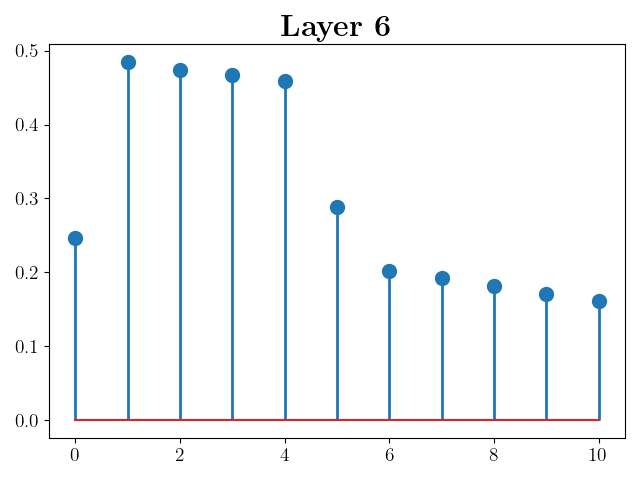} &
\includegraphics[trim={0 0 0 0},clip, width=0.23\linewidth]{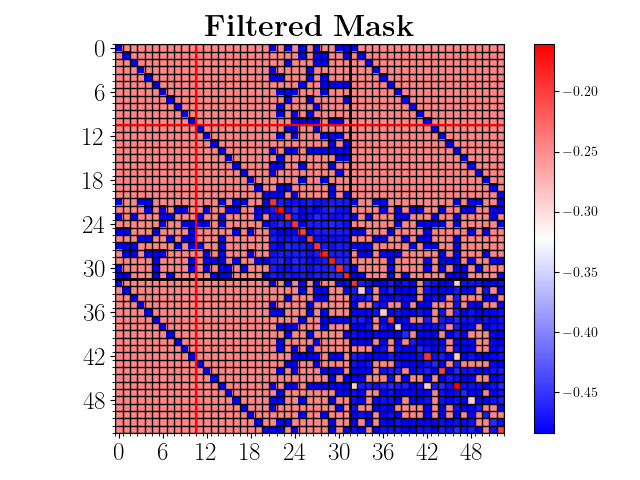} &
\includegraphics[trim={0 0 0 0},clip, width=0.23\linewidth]{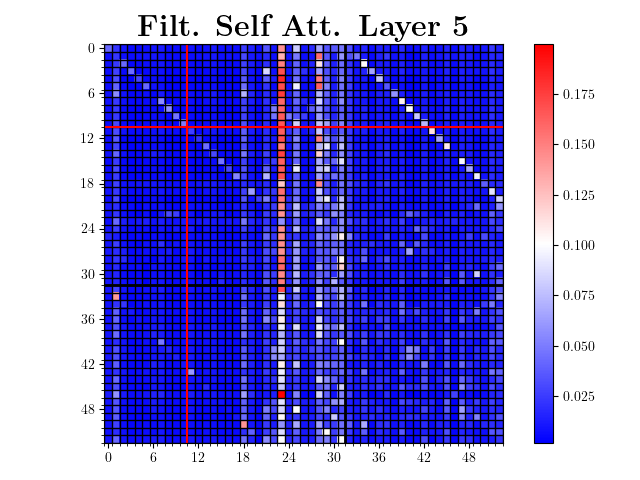} \\
      \end{tabular}
      }
  \caption{
  \textcolor{black}{
For a $N=6$ layers DC-ECCT (first and second row) and a (32,11) code: (a) self-attention layer, (b) connectivity mapping $\psi_{\gamma}$, (c) the corresponding filtered mask $\psi_{\gamma}\big(g(H_{\Omega})\big)$ (d) the obtained soft masked self-attention. The self-attention maps have been averaged over the heads dimension.}
}\label{fig:app_analysis_filters}
\end{figure*}

Figure \ref{fig:app_analysis_filters} depicts BER plots comparing the performance of the baselines and the proposed method for several codes.

\section{Ablation Study}
\label{appendix:ablation}
In Figure \ref{tab:ablation} we present an ablation study of the different elements of the proposed framework.

In order to emphasize the importance of a mask induced by the code we provide two ablation studies.
Regarding the \emph{masking} procedure itself, we first provide the performance of a fully trainable mask, such that the self-attention is multiplied similarly as in Eq \ref{eq:e2e_mask} while the mask is a parameter matrix trained independently of the code.
Regarding the training via the mask we show the performance of the proposed framework where the gradients originated from the mask are stopped.

In order to emphasize the importance of the sampling of the original message $m$ we provide results using random sampling of $m$ (i.e., not $m=1$). We note here that using $m=0$ (i.e., $G$ is not taken into account for the optimization), as done in almost all the existing neural decoding methods, just completely fails the training.

Finally, even though there exists no known alternative, we provide an interesting insight into the gradient computation of the modulo operator.
We apply the STE \cite{bengio2013estimating} on the results of a regular matrix-vector multiplication. Given the modulo function $g$, we then have $g(x)=x \text{mod}2$ and $\frac{\partial g(x)}{\partial x} = \mathbbm{1}_{|x|\leq 1}$. Since this approach is less theoretically founded than the proposed polarization (especially in our scenario where the modulo operation is performed over $\mathbb{N}$), it is interesting to see its overall training seems to bring enhanced performance.

{\color{black} We note here that the quality of the learned code can certainly be improved if better trained on larger noise ranges (i.e., 
 range) and batch size, depending on available computing resources.}

\begin{table}[t]
    \centering
    \caption{    Ablation analysis on two codes for the $N=2,d-32$ architecture. For fairness, the same setting is used in all experiment: clamping $\Omega$ s.t. $|\Omega|\leq 1$ and we stop the training of $\Omega$ after 800 iterations among 1000. 
	$\Omega_{0}$ is randomely sampled. We show the performance of the proposed method (Our), the performance using a fully trainable mask independently of the code (Mask V2), and the performance of using the STE for the modulo calculations instead of the polarization approximation. 
	We show the performance when we stop the gradient from backpropagating via the mask (Mask Stop Gradient) as well as the performance of using random $m$ instead of all ones.}
    \label{tab:ablation}
    \smallskip
    \begin{tabular}{lcccccc}
    \toprule
        Code & \multicolumn{3}{c}{(31,16)} & \multicolumn{3}{c}{(32,11)}\\
        \cmidrule(lr){2-4}
		\cmidrule(lr){5-7}
         $E_{b}/N_{0}$ & 4 & 5 & 6 & 4 & 5 & 6 \\ 
 \midrule   
{Our}    & 5.16      & 6.51	        & 8.19 		& 4.40      & 5.54	        & 7.04		\\
Mask V2 	    & 4.83      & 6.18	        & 7.85 		& 4.04      & 5.05	        & 6.28		\\
STE	       		& 5.11      & 6.53	        & 8.20 		& 4.63      & 5.89	        & 7.39		\\
Mask S.G.	    & 4.54      & 5.81	        & 7.50 		& 4.26      & 5.39	        & 6.79		\\
Random $m$  	& 5.04      & 6.44	     	& 8.13 		& 4.38      & 5.50	        & 6.84		\\
\bottomrule
	\end{tabular}
\end{table} 


\section{BER Visualization}
\label{appendix:ber_vis}
\begin{figure*}[h]
\centering
\resizebox{0.885\textwidth}{!}
{
    \noindent  \begin{tabular}{@{}ccc@{}}
        \includegraphics[trim={0 0 0 0},clip, width=0.23\linewidth]{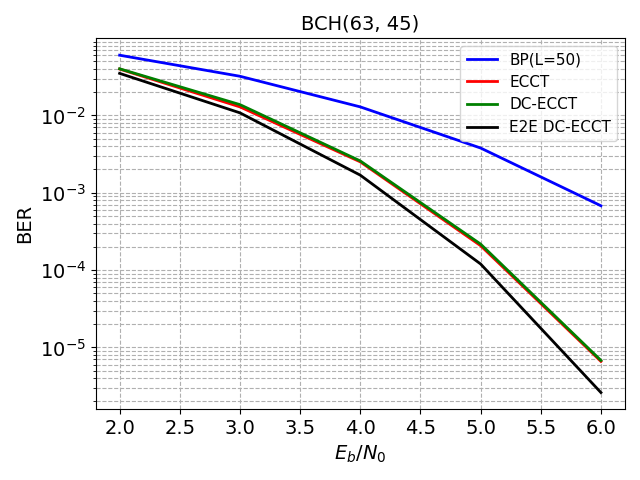} &
        \includegraphics[trim={0 0 0 0},clip, width=0.23\linewidth]{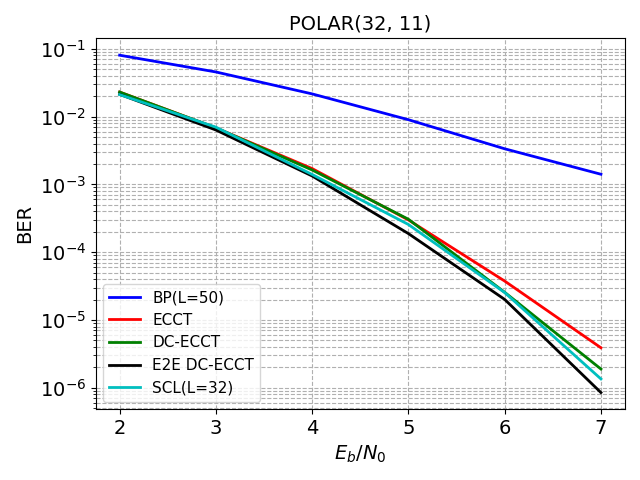} &
        \includegraphics[trim={0 0 0 0},clip, width=0.23\linewidth]{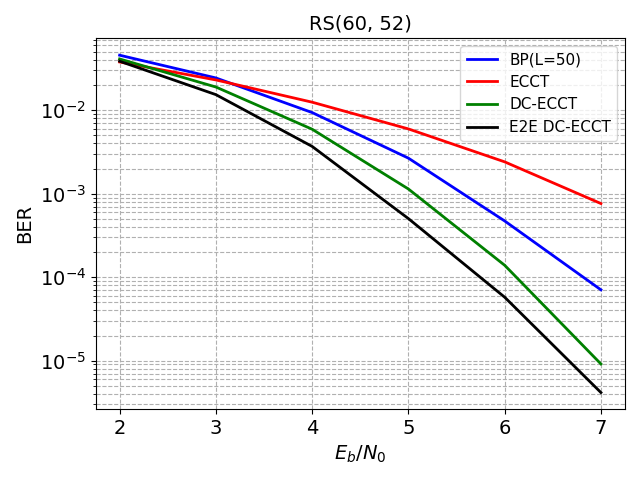} \\
    \end{tabular}
}
\caption{
    \textcolor{black}{
    BER plots comparing the performance of the baselines and the proposed method for various $E_{b}/N_{0}$ values.
    }
}\label{fig:app_ber_plots}
\end{figure*}

Figure \ref{fig:app_ber_plots} depicts classical BER plots for three of the tested codes.




\end{document}